\documentclass[twocolumn]{aastex6}

\usepackage{amssymb,amsmath,graphicx,latexsym}
\usepackage{bm}
\usepackage{epsfig}
\usepackage{subfigure}

\shorttitle{Resurrecting the Power-law, Intermediate, and Logamediate Inflations ...}
\shortauthors{Amani et al.}
\begin{document}


\title{Resurrecting the Power-law, Intermediate, and Logamediate Inflations in  the DBI Scenario with Constant Sound Speed}

%% Use \author, \affil, and the \and command to format author and affiliation 
%% information.  If done correctly the peer review system will be able to
%% automatically put the author and affiliation information from the manuscript
%% and save the corresponding author the trouble of entering it by hand.
%%
%% The \affil should be used to document primary affiliations and the
%% \altaffil should be used for secondary affiliations, titles, or email.

%% Authors with the same affiliation can be grouped in a single
\author{Roonak Amani$^1$, Kazem Rezazadeh$^1$, Asrin Abdolmaleki$^2$, and Kayoomars Karami$^1$}
\affil{$^1$Department of Physics, University of Kurdistan, Pasdaran Street, P.O. Box 66177-15175, Sanandaj, Iran; rezazadeh86@gmail.com\\
$^2$Research Institute for Astronomy and Astrophysics of Maragha (RIAAM), P.O. Box 55134-441, Maragha, Iran}

%% AASTeX 6.0 supports the ability to suppress the names and affiliations
%% of some authors and displaying them under a "collaboration" banner to
%% minimize the amount of author information that to be printed.  This 
%% should be reserved for articles with an extreme number of authors.  
%% The necessary command are \AuthorCallLimit and \collaborationName.
%% An \AuthorCallLimit=2 call prior to the author list will only show
%% the authors in the first two \author calls.  The \collaborationName
%% defines the collaboration identifier.  Commented examples below.

%\AuthorCallLimit=1
%% Will only show Schwarz & Muench since Schwarz and Muench
%% are in the same \author call. 
%\collaborationName{Friends of AASTeX}
%% will print "The AAS collaboration" after the shortened author list.
%% Note that all the \altaffil information will still be shown so it
%% has to be manually commented out if you do not want it shown.
%%
%% Note that all of these author will be shown in the published article.
%% This feature is meant to be used prior to acceptance to make the
%% front end of a long author article more manageable.

\begin{abstract}

We investigate the power-law, intermediate, and logamediate inflationary models in the framework of DBI non-canonical scalar field with constant sound speed. In the DBI setting, we first represent the power spectrum of both scalar density and tensor gravitational perturbations. Then, we derive different inflationary observables including the scalar spectral index $n_s$, the running of the scalar spectral index $dn_s/d\ln k$, and the tensor-to-scalar ratio $r$. We show that the 95\% CL constraint of the Planck 2015 T+E data on the non-Gaussianity parameter $f_{{\rm NL}}^{{\rm DBI}}$ leads to the sound speed bound $c_{s}\geq0.087$ in the DBI inflation. Moreover, our results imply that, although the predictions of the power-law, intermediate, and logamediate inflations in the standard canonical framework ($c_s=1$) are not consistent with the Planck 2015 data, in the DBI scenario with constant sound speed $c_s<1$, the result of the $r-n_s$ diagram for these models can lie inside the 68\% CL region favored by the Planck 2015 TT,TE,EE+lowP data. We also specify the parameter space of the power-law, intermediate, and logamediate inflations for which our models are compatible with the 68\% or 95\% CL regions of the Planck 2015 TT,TE,EE+lowP data. Using the allowed ranges of the parameter space of the intermediate and logamediate inflationary models, we estimate the running of the scalar spectral index and find that it is compatible with the 95\% CL constraint from the Planck 2015 TT,TE,EE+lowP data.

\end{abstract}




\section{Introduction}
\label{section:introduction}

Inflation is a central part of modern cosmology. In this scenario, it is assumed that a fast accelerated expansion has taken place in a short period of time before the radiation-dominated era \citep{Guth1981, Albrecht1982, Linde1982, Linde1983}. Consideration of this accelerated expansion can successfully resolve the well-known problems of classical cosmology  \citep{Guth1981, Albrecht1982, Linde1982, Linde1983}. Besides, the perturbations created during inflation can be regarded as the origin of the large-scale structure (LSS) formation as well as the anisotropies of the cosmic microwave background (CMB) radiation \citep{Mukhanov1981, Hawking1982, Guth1982, Starobinsky1982}. Indeed, the general prediction of the inflationary scenario that the primordial perturbations should be adiabatic and nearly scale-invariant is in well agreement with the recent observational data provided by the Planck satellite \citep{Planck2015, Planck2015non-Gaussianity}.

Despite of its achievements, the inflationary paradigm suffers from fundamental ambiguities. In particular, so far we do not know what the nature of the scalar field driving inflation (inflaton) is, what shape the inflaton potential possesses, at which energy scale the inflationary era of our universe occurred, and what gravitational theory is valid at that era \citep{Baumann2009}. In addition, the conventional models of inflation suffer from the eta problem that challenges the basic principles of them \citep{Copeland1994}. There exist no convincing solutions to these problems in the context of the standard inflationary framework based on a single canonical scalar field in the Einstein gravity \citep{Baumann2009, Martin2014}. However, so far remarkable attempts have been made to solve these drawbacks in the alternative models to the standard canonical inflation. One important class of these alternative models is the Dirac-Born-Infeld (DBI) inflation, which has robust motivations from string theory \citep{Alishahiha2004, Silverstein2004}. In the DBI inflation, the role of the inflaton is played by the radial coordinate of a D3-brane moving in a warped region (throat) of a compactification space \citep{Alishahiha2004, Silverstein2004}. From the physical standpoint, this may be interpreted as the impact of the higher dimensions on our four-dimensional spacetime, which leads to the appearance of the DBI Lagrangian in the effective action through the dimensional reduction. Hence, the DBI framework proposes a convincing candidate for the inflaton field. Furthermore, by investigating the dynamics of the inflaton on the throat of the compactification space, some suggestions have been presented for the inflaton potential that make it possible for us to circumvent the eta problem \citep{Kachru2003, Firouzjahi2005, Shandera2006, Baumann2007, Baumann2008, Baumann2009-2, Krause2008, Easson2009}.

From the theoretical point of view, the DBI inflation can be included in the class of the non-canonical inflationary models in which the kinetic term in the action is different from the canonical one \citep{Armendariz-Picon1999, Garriga1999}. The outstanding feature of the non-canonical inflationary models is the fact that in these models, the propagation speed of the density perturbations, so-called the ``sound speed,'' can be different from the light speed. Due to this property, the value of the tensor-to-scalar ratio is generally reduced in this class of models  relative to the standard canonical inflationary setting \citep{Unnikrishnan2012}, which supplies a better consistency with the Planck 2015 observational data \citep{Planck2015}. In addition, because of this property, the non-canonical inflationary models are capable of providing rather large values for the primordial non-Gaussianity \citep{Alishahiha2004, Chen2007, Li2008, Tolley2010}, and this feature may be used to discriminate between the canonical and non-canonical inflationary models by increasingly precise measurements in the future.

Inflation with the DBI scalar field has been extensively studied in the literature \citep{Kachru2003,Alishahiha2004,Silverstein2004,Chen2005,Chen2005-2,Chen2005-3,Firouzjahi2005,Shandera2006,Baumann2007,Chen2007,Peiris2007,Spalinski2007,Baumann2008, Gauthier2008,Krause2008,Li2008,Spalinski2008,Baumann2009-2,Bessada2009,Easson2009,Tzirakis2009,Tolley2010,Cai2011,van-de-Bruck2011,Miranda2012,Tsujikawa2013,Nazavari2016,Rasouli2017}. The basic concepts of this model have been introduced in \citet{Silverstein2004}, where the authors have analyzed the motion of a rolling scalar field expressly in the strong coupling regime of the field theory and have developed their study to cosmological contexts based on the minimal coupling of this kind of field theory to four-dimensional gravity. In \citet{Alishahiha2004}, the authors have studied the perturbations in the DBI inflation to the nonlinear order and have shown that the primordial non-Gaussianity parameter has a direct relation with the Lorentz factor (or, equivalently, with the sound speed) of the model, and consequently this model is capable of providing large values for the primordial non-Gaussianity. Further studies done later verified these conclusions \citep{Chen2007, Li2008, Tolley2010}. In \citet{Kachru2003, Firouzjahi2005, Shandera2006, Baumann2007, Baumann2008, Baumann2009-2, Krause2008}, and \citet{Easson2009}, some attempts have been made to reconstruct the inflationary potentials in the DBI setting to ameliorate the eta problem in describing the dynamics of the early universe. The Hamilton--Jacobi formalism in the context of DBI scalar field theories was introduced in \citet{Silverstein2004}, and then was discussed with more details in \citet{Spalinski2007}. Different DBI inflationary models have been investigated in the references \citet{Gauthier2008, Peiris2007, Bessada2009, Tzirakis2009, Miranda2012, Tsujikawa2013}, and \citet{Nazavari2016}, and validity of their predictions has been appraised with the CMB data. In \citet{van-de-Bruck2011}, the DBI inflation has developed in the context of a scalar-tensor theory of gravity. Finally, in \citet{Cai2011}, and \citet{Rasouli2017}, the DBI inflation has been studied in the warm inflationary scenario in which during inflation the inflaton energy density is transformed to the radiation energy density constantly, so that the universe can transit to the radiation dominated era without resorting to any additional reheating processes.

In the present work, we study the power-law \citep{Lucchin1985, Halliwell1987, Yokoyama1988, Liddle1989}, intermediate \citep{Barrow1990, Barrow1993, Barrow2006, Barrow2007}, and logamediate \citep{Barrow2007} scale factors in the DBI inflation with constant sound speed. These scale factors have special importance in cosmological contexts because they appear as a class of possible indefinite solutions from imposing weak general conditions on cosmological models. In \cite{Barrow1996}, it has been discussed that we can find eight conceivable asymptotic cosmological solutions, of which three give rise to non-inflationary expansions. Three others result in the power-law ($a(t)\propto t^{q}$ where $q>1$), de Sitter ($a(t)\propto \exp{(Ht)}$ where $H$ is constant), and intermediate ($a(t)\propto\exp{\big(At^\lambda\big)}$ where $A>0$ and $0<\lambda<1$) inflationary expansions. The remaining two inflationary expansions have asymptotic behaviors in logamediate form. Although these scale factors have remarkable cosmological significance, they are not consistent with the Planck 2015 observational results \citep{Planck2015} in the standard canonical inflationary scenario ($c_s=1$), as it has been explicitly shown in \citet{Rezazadeh2015, Rezazadeh2016, Rezazadeh2017}. In the present paper, we aim to refine the power-law, intermediate, and logamediate inflationary models in light of the current observational results by considering them in the DBI scalar field setting. Motivated by \citet{Spalinski2008}, and \citet{Tsujikawa2013}, we assume in our work that the sound speed is constant during inflation. We examine the viability of each scale factor in light of the Planck 2015 results \citep{Planck2015} and also present some observational constraints on them.

The outline of this paper is as follows. In Sec. \ref{section:DBI}, we review briefly the basic equations governing the background cosmology in the DBI inflation. In addition, we also represent the power spectra of the primordial scalar and tensor perturbations. Then, in Secs. \ref{section:power-law}, \ref{section:intermediate}, and \ref{section:logamediate}, we respectively investigate the power-law, intermediate, and logamediate DBI inflationary models with constant sound speed. Finally, in Sec. \ref{section:conclusions}, we summarize our conclusions.

\section{DBI inflation}
\label{section:DBI}

Within the framework of Einstein's gravity, the action of non-canonical scalar field models is given by \citep{Armendariz-Picon1999, Garriga1999}
\begin{equation}
\label{S}
S=\int d^{4}x\sqrt{-g}\left[\frac{1}{2}R+\mathcal{L}(X,\phi)\right],
\end{equation}
where $g$ and $R$ are the determinant of the background metric and the Ricci scalar, respectively. Throughout this paper, we work in the units where the Planck reduced mass is taken equal to unity, $M_{P}=(8\pi G)^{-1/2}$=1, for the sake of simplicity. In the present work, we concentrate on the DBI non-canonical scalar field with the Lagrangian density \citep{Alishahiha2004, Silverstein2004}
\begin{equation}
\label{{mathcal}{L}}
\mathcal{L}(X,\phi)\equiv f^{-1}(\phi)\left[1-\sqrt{1-2f(\phi)X}\right]-V(\phi),
\end{equation}
where $f(\phi)$ is the warp factor, while $X\equiv -g^{\mu\nu}\phi_{,\mu}\phi_{,\nu}/2$ and $V(\phi)$ are the canonical kinetic term and potential of the DBI scalar field $\phi$, respectively.

For the DBI Lagrangian (\ref{{mathcal}{L}}), the energy density and pressure of the non-canonical scalar field $\phi$ are given by
\begin{eqnarray}
\label{{rho}_{phi}}
\rho_\phi &\equiv& 2X\mathcal{L}_{,\rm X} -\mathcal{L}=\frac{\gamma-1}{f(\phi)}+V(\phi),
\\
\label{p_{phi}}
p_\phi &\equiv& \mathcal{L}=\frac{\gamma-1}{\gamma f(\phi)}-V(\phi),
\end{eqnarray}
where the subscript ``$,X$'' indicates the partial derivative with respect to $X$. Moreover,
\begin{equation}
\label{{gamma}}
\gamma\equiv\frac{1}{\sqrt{1-2f(\phi)X}},
\end{equation}
is the parameter determining the relativistic limit of brane motion in a warped background. Note that the parameter $\gamma$ is defined in analogy of the Lorentz factor appearing in the relativistic particle dynamics.

The sound speed is generally defined by
\begin{equation}
\label{c_s,definition}
c^2_s \equiv \frac{p_{\phi,{\rm X}}}{\rho_{\phi,{\rm X}}},
\end{equation}
which specifies the propagation speed of the inflaton fluctuations $\delta\phi$ relative to the homogeneous background. In order for the sound speed to be physical, it should be real and subluminal, $0<c_{s}^{2}\leq1$ \citep{Franche2010}. Applying Equations (\ref{{rho}_{phi}})--(\ref{{gamma}}), the sound speed in DBI inflation takes the form
\begin{equation}
\label{c_s}
c_{s}=\sqrt{1-2f(\phi)X}=\frac{1}{\gamma}.
\end{equation}
%Substituting $c_s$ from Equation (\ref{c_s}) into (\ref{{mathcal}{L}}) one can rewrite the DBI Lagrangian in a simple form as
%\begin{equation}
%\label{{mathcal}{L},c_s}
%\mathcal{L}(X,\phi)=\left(\frac{2}{1+c_s}\right)X-V(\phi),
%\end{equation}
%which explicitly shows that for $c_s=1$, the DBI Lagrangian is converted to the standard canonical Lagrangian $\mathcal{L}(X,\phi)=X-V(\phi)$.

For the homogeneous and isotropic background space-time, we consider the spatially flat Friedmann-Robertson-Walker (FRW) metric. Therefore, varying the action (\ref{S}) with respect to the metric yields the Friedmann equations in the DBI inflation as follows:
\begin{eqnarray}
\label{H,{rho}_{phi}}
H^{2} &=& \frac{1}{3}\rho_{\phi},
\\
\label{{dot}{H},{rho}_{phi},p_{phi}}
\dot{H} &=& -\frac{1}{2}\left(\rho_{\phi}+p_{\phi}\right).
\end{eqnarray}
Here, $H=\dot{a}/a$ is the Hubble parameter, and $a$ is the scale factor $a$ of the universe.  The dot denotes the derivative with respect to the cosmic time $t$. Considering the FRW metric, the canonical kinetic term is simplified to $X=\dot{\phi}^2/2$.

Taking the variation of action (\ref{S}) with respect to $\phi$ gives the equation of motion of the scalar field as
\begin{equation}
\label{{ddot}{{phi}}}
\ddot{\phi}+\frac{3H}{\gamma^{2}}\dot{\phi}+\frac{V'(\phi)}{\gamma^{3}}+\frac{f'(\phi)(\gamma+2)(\gamma-1)}{2f(\phi)\gamma(\gamma+1)}\dot{\phi}^{2}=0,
\end{equation}
where the prime denotes the derivative with respect to $\phi$. For the case of $\gamma=1$,  this equation recovers the well-known equation of motion in the standard model of inflation. Note that Equation (\ref{{ddot}{{phi}}}) can also be obtained from the continuity equation
\begin{equation}
\label{{dot}{{rho}}_{phi}}
\dot{\rho}_{\phi}+3H(\rho_{\phi}+p_{\phi})=0,
\end{equation}
where $\rho_{\phi}$ and $p_{\phi}$ are given by Equations (\ref{{rho}_{phi}}) and (\ref{p_{phi}}), respectively.

Substituting Equations (\ref{{rho}_{phi}}) and (\ref{p_{phi}}) into (\ref{{dot}{H},{rho}_{phi},p_{phi}}), and using Equation (\ref{{gamma}}), one can easily obtain
\begin{equation}
\label{{dot}{{phi}},H'}
\dot{\phi}=-\frac{2}{\gamma}H'(\phi),
\end{equation}
which is the same as that obtained for the DBI inflationary model in the Hamilton--Jacobi formalism \citep{Spalinski2007}.

In the study of inflation, it is convenient to express the extent of the universe expansion in terms of the $e$-fold number
\begin{equation}
\label{N}
N\equiv\ln\frac{a_{e}}{a}.
\end{equation}
The above definition is equivalent to
\begin{equation}
\label{dN}
dN=-Hdt.
\end{equation}

In the DBI inflationary framework, it is useful to define the Hubble slow-roll parameters as follows:
\begin{eqnarray}
\label{{varepsilon}_1}
\varepsilon_{1} &\equiv& -\frac{\dot{H}}{H^{2}},
\\
\label{{varepsilon}_{i+1}}
\varepsilon_{i+1} &\equiv& \frac{\dot{\varepsilon}_{i}}{H\varepsilon_{i}},\qquad(i\geq1).
\end{eqnarray}
Furthermore, it is convenient to define sound slow-roll parameters in the following forms:
\begin{eqnarray}
\label{{varepsilon}_{s1}}
\varepsilon_{s1} &\equiv& \frac{\dot{c}_{s}}{Hc_{s}},
\\
\label{{varepsilon}_{s(i+1)}}
\varepsilon_{s(i+1)} &\equiv& \frac{\dot{\varepsilon}_{si}}{H\varepsilon_{si}},\qquad(i\geq1).
\end{eqnarray}
During inflation, the Hubble and sound slow-roll parameters are usually much less than unity, and considering these conditions is known as the slow-roll approximation.

It is straightforward to show that using Equations (\ref{{rho}_{phi}}), (\ref{c_s}), (\ref{{dot}{{phi}},H'}), and (\ref{{varepsilon}_1}), the Friedmann equation (\ref{H,{rho}_{phi}}) can be written as
\begin{equation}
 \label{H,V,{varepsilon}_1}
 \frac{V(\phi)}{3H^{2}}+\frac{2}{3}\frac{\varepsilon_{1}}{1+c_{s}}=1.
\end{equation}
Since in the slow-roll regime, $\varepsilon_{1}\ll1$, we therefore can ignore the second term in the left--hand side of the above equation versus the first one. Consequently, in the DBI inflation, the first Friedmann equation in the slow-roll approximation takes the form \citep{Chen2007, Miranda2012}
\begin{equation}
 \label{H,V}
 H^{2}\simeq\frac{1}{3}V(\phi),
\end{equation}
which is same as that obtained in the standard canonical inflation. In addition, in the slow-roll approximation, the first term on the left--hand side of Equation (\ref{{ddot}{{phi}}}) can be neglected in front of the next terms, and therefore using Equations (\ref{{gamma}}) and (\ref{c_s}), we find \citep{Miranda2012}
\begin{equation}
\label{{dot}{{phi}}}
3H\dot{\phi}+c_{s}V'\simeq0.
\end{equation}

In this paper, our main goal is to compare the results of some DBI inflationary models with the Planck 2015 CMB data \citep{Planck2015}. Therefore, it is useful to mention how the Planck Collaboration \citep{Planck2015} applies the CMB anisotropy data to constrain the inflationary observables. The Planck Collaboration \citep{Planck2015} connects the CMB angular power spectra to the scalar and tensor primordial power spectra via the transfer
functions $\Delta_{l,\mathcal{A}}^{s}(k)$ and $\Delta_{l,\mathcal{A}}^{t}(k)$ as follows:
\begin{align}
 \label{C_l^{{mathcal}{A}{mathcal}{B},s}}
 C_{l}^{\mathcal{A}\mathcal{B},s} &=\int_{0}^{\infty}\frac{dk}{k}
 \Delta_{l,\mathcal{A}}^{s}(k)
 \Delta_{l,\mathcal{B}}^{s}(k)\mathcal{P}_{s}(k),
 \\
 \label{C_l^{{mathcal}{A}{mathcal}{B},t}}
 C_{l}^{\mathcal{A}\mathcal{B},t} &=\int_{0}^{\infty}\frac{dk}{k}
 \Delta_{l,\mathcal{A}}^{t}(k)
 \Delta_{l,\mathcal{B}}^{t}(k)\mathcal{P}_{t}(k),
\end{align}
where $\mathcal{A},\mathcal{B}=T,E,B$ refer to temperature mode T and polarization modes E and B. The transfer functions $\Delta_{l,\mathcal{A}}^{s}(k)$ and $\Delta_{l,\mathcal{A}}^{t}(k)$ specify the evolution of the perturbations after their re-entry into the Hubble horizon. These functions generally have to be computed numerically using the Boltzmann codes such as CMBFAST \citep{Seljak1996} or CAMB \citep{Lewis2000}, which depend on the parameters of the background cosmological model. The Planck Collaboration \citep{Planck2015} assumes that the background dynamics of the universe after inflation is followed by the standard cosmological model $\Lambda$CDM. In addition, the Planck team \citep{Planck2015} parameterizes phenomenologically the primordial power spectra of the scalar and tensor perturbations, respectively, as follows:
\begin{align}
 &\mathcal{P}_{s}(k) = \mathcal{P}_{s}\left(k_{*}\right)
 \nonumber
 \\
 &\times\left(\frac{k}{k_{*}}\right)^{n_{s}-1+\frac{1}{2}dn_{s}/d\ln k\ln(k/k_{*})+\frac{1}{6}\frac{d^{2}n_{s}}{d\ln k^{2}}(\ln(k/k_{*}))^{2}+...},
 \label{{mathcal}{P}_s(k)}
 \\
 &\mathcal{P}_{t}(k) = \mathcal{P}_{t}\left(k_{*}\right)\left(\frac{k}{k_{*}}\right)^{n_{t}+\frac{1}{2}dn_{t}/d\ln k\ln(k/k_{*})+...},
 \label{{mathcal}{P}_t(k)}
\end{align}
where $n_s$ ($n_t$), $dn_s/d\ln k$ ($dn_t/d\ln k$), and $d^2 n_s /d(\ln k)^2$ are the scalar (tensor) spectral index, the running of the scalar (tensor) spectral index, and the running of the running of the scalar spectral index, respectively. The ratio of the tensor power spectrum to the scalar one is denoted by $r$, and that is a crucial inflationary observable to discriminate between inflationary models. All of these observables are evaluated at the pivot scale denoted by $k_*$. Note that definitions (\ref{{mathcal}{P}_s(k)}) and (\ref{{mathcal}{P}_t(k)}) are completely independent of the inflationary model, and consequently the Planck 2015 results \citep{Planck2015} are applicable for each model in which the subsequent evolution of the universe after inflation is described by the $\Lambda$CDM model. Throughout this paper, we focus on the Planck 2015 data set fitted on the base of the model $\Lambda$CDM+$r$+$dn_s/d\ln k$. In this model, it is supposed that a quasi-de Sitter expansion has happened during inflation. Note that a quasi-de Sitter expansion generally leads to almost scale-invariant power spectra for the primordial perturbations, which is in good agreement with the recent observational data. In our work, we also assume that the perturbations generated during the slow-roll inflation are adiabatic, so that we can neglect the entropy generation during this period. We further assume after inflation, the dynamics of the background cosmology is followed by the $\Lambda$CDM model. In the model $\Lambda$CDM+$r$+$dn_s/d\ln k$, only the amplitude of the tensor perturbations $\mathcal{P}_{t}(k_{*})$ is taken into account, and the scale-dependence of these perturbations, which is expressed by $n_t$, is supposed to be negligible. But regarding the scalar perturbations, not only their scale-dependence which is measured by $n_s$, is included, but also the scale-dependence of the quantity $n_s$, which is given by $dn_s/d\ln k$, is taken into account. Here, we use all of assumptions taken by the Planck Collaboration \citep{Planck2015} and apply their observational constraints on inflation to examine the viability of our DBI inflationary model.

Now, we turn to obtaining the equations of the inflationary observables in the DBI inflationary scenario. For this purpose, we first note that the power spectrum of the scalar perturbations in this model is given by \citep{Garriga1999}
\begin{align}
 \label{{mathcal}{P}_s}
 \mathcal{P}_{s}=\left.\frac{H^{2}}{8\pi^{2}c_{s}\varepsilon_{1}}\right|_{c_{s}k=aH}.
\end{align}
This expression should be evaluated at the time of sound horizon exit at which $c_{s}k=aH$. The recent value reported by the Planck Collaboration for the scalar perturbation amplitude at the horizon crossing of the pivot scale $k_{*}=0.05\,{\rm Mpc}^{{\rm -1}}$ is $\ln\left[10^{10}{\cal P}_{s}\left(k_{*}\right)\right]=3.094\pm0.034$ \citep[Planck 2015 TT,TE,EE+lowP data;][]{Planck2015}. In order to specify the scale-dependence of the scalar perturbations, it is conventional to define the scalar spectral index as
\begin{equation}
\label{n_s,definition}
n_{s}-1\equiv\frac{d\ln\mathcal{P}_{s}}{d\ln k}.
\end{equation}
The reported value by the Planck 2015 collaboration for the scalar spectral index is $ n_{s}=0.9644 \pm 0.0049 $ \citep[68\% CL, Planck 2015 TT,TE,EE+lowP;][]{Planck2015}. In the slow-roll regime, the Hubble parameter $H$ and the sound speed $c_s$ vary much more slowly than the scale factor $a$ of the universe \citep{Garriga1999}, and hence the relation $c_{s}k=aH$ leads to
\begin{equation}
\label{d{ln}k}
d\ln k\approx Hdt=-dN.
\end{equation}
Applying Equations (\ref{{varepsilon}_1})--(\ref{{varepsilon}_{s1}}) and (\ref{{mathcal}{P}_s})--(\ref{d{ln}k}), we can obtain a relation for the scalar spectral index in the non-canonical inflationary framework as
\begin{equation}
\label{n_s}
n_{s}=1-2\varepsilon_{1}-\varepsilon_{2}-\varepsilon_{s1}.
\end{equation}
Furthermore, with the help of Equations (\ref{{varepsilon}_1})--(\ref{{varepsilon}_{s(i+1)}}), and (\ref{n_s}), we can obtain the running of the scalar spectral index as
\begin{equation}
\label{dn_s/d{ln}k}
\frac{dn_{s}}{d\ln k}=-2\varepsilon_{1}\varepsilon_{2}-\varepsilon_{2}\varepsilon_{3}-\varepsilon_{s1}\varepsilon_{s2}.
\end{equation}
The CMB constraint on this observable is $dn_{s}/d\ln k=-{\rm 0}.0085\pm{\rm 0}.0076$ \citep[68\% CL, Planck 2015 TT, TE, EE+lowP;][]{Planck2015}.

Now, we concentrate on the tensor perturbations in the setting of DBI non-canonical inflation. The tensor power spectrum of this scenario is given by \citep{Garriga1999}
\begin{equation}
 \label{{mathcal}{P}_t}
 \mathcal{P}_{t}=\left.\frac{2H^{2}}{\pi^{2}}\right|_{k=aH}.
\end{equation}
It should be noted that, in contrary to the scalar perturbations that freeze when they leave the sound horizon ($c_{s}k=aH$), the tensor perturbations freeze when they cross the Hubble horizon ($k=aH$) outward of it. However, the difference between the sound horizon exit and the Hubble horizon exit is negligible to the lowest order in the slow-roll parameters \citep{Garriga1999, Unnikrishnan2012}. One should note that the tensor power spectrum (\ref{{mathcal}{P}_t}) in the non-canonical inflationary setting is the same as that obtained in the standard model of canonical inflation because the tensor modes (or gravitational waves) depend on the gravitational part of the action (\ref{S}) that is identical in both of these models that are based on Einstein's gravity.

The scale-dependence of the tensor perturbations is measured by the tensor spectral index defined as
\begin{equation}
\label{n_t,definition}
n_{t}\equiv\frac{d\ln\mathcal{P}_{t}}{d\ln k}.
\end{equation}
No precise measurement for the tensor spectral index $n_t$ currently exists, so we have to wait for the future observations \citep{Simard2015}. Using Equations (\ref{{varepsilon}_1}), (\ref{d{ln}k}), and (\ref{{mathcal}{P}_t}), the tensor spectral index (\ref{n_t,definition}) reads
\begin{equation}
\label{n_t}
n_t=-2\varepsilon_1.
\end{equation}

As mentioned above, the tensor-to-scalar ratio is defined as
\begin{equation}
\label{r,definition}
r\equiv\frac{\mathcal{P}_{t}}{\mathcal{P}_{s}}.
\end{equation}
The Planck 2015 data have provided the upper limit on tensor-to-scalar ratio as $r<0.149$ \citep[95\% CL, Planck 2015 TT,TE,EE+lowP;][]{Planck2015}, which is a central criterion to discriminate between inflationary models. For the DBI inflation, using Equations (\ref{{mathcal}{P}_s}) and (\ref{{mathcal}{P}_t}), it is straightforward to show that the tensor-to-scalar ratio (\ref{r,definition}) takes the form
\begin{equation}
\label{r}
r=16c_{s}\varepsilon_{1},
\end{equation}
where we have neglected the difference between the sound horizon exit and the Hubble horizon exit in the slow-roll approximation \citep{Garriga1999, Unnikrishnan2012}.

Combining Equations (\ref{n_t}) and (\ref{r}), it is easy to get
\begin{equation}
 \label{r,n_t}
 r=-8c_s n_t,
\end{equation}
which is known as the consistency relation of the DBI inflation. This differs from the standard consistency relation $r=-8n_t$ in the canonical inflation. However, as it was explained previously, the Planck 2015 constraints on inflation that we will consider in the present paper are applicable for all slow-roll inflationary models, independent of their consistency relations. Of course, if one considers the Planck bounds on $r$ and applies the consistency relation of the model, one can find some constraints on $n_t$ that are obviously model-dependent.

One other important observable is the primordial non-Gaussianity that can be generated during inflation. It can be used as a powerful criterion to discriminate between different inflationary scenarios \citep{Babich2004}. There are different types of non-Gaussianity parameters containing the equilateral, squeezed, and orthogonal shapes that arise respectively in models with higher-derivative interactions resulting in non-trivial speeds of sound \citep{Alishahiha2004, Chen2007, Li2008, Tolley2010}, multi-fields inflation \citep{Bartolo2002, Sasaki2008}, and models with non-standard initial states \citep{Chen2007, Holman2008}. In our case, we deal with a single field inflation, and so the signal of primordial non-Gaussianity is peaked at the equilateral triangle configuration.

For the DBI models of inflation, the Planck Collaboration \citep{Planck2015non-Gaussianity}, following \citet{Alishahiha2004}, considers phenomenologically the bispectrum of the primordial perturbations as
\begin{align}
 & B_{\Phi}^{\mathrm{DBI}}(k_{1},k_{2},k_{3})= \frac{6A^{2}f_{\mathrm{NL}}^{\mathrm{DBI}}}{\left(k_{1}k_{2}k_{3}\right)^{3}}\frac{\left(-3/7\right)}{(k_{1}+k_{2}+k_{3})^{2}}
 \nonumber
 \\
 &\times\Big\{ \underset{i}{\sum}k_{i}^{5}+\underset{i\neq j}{\sum}\left[2k_{i}^{4}k_{j}-3k_{i}^{3}k_{j}^{2}\right]
 \nonumber
 \\
 &+\underset{i\neq j\neq l}{\sum}\left[k_{i}^{3}k_{j}k_{l}-4k_{i}^{2}k_{j}^{2}k_{l}\right]\Big\},
 \label{B_{Phi}}
\end{align}
which is close to the equilateral shape. Here, the subscript $\Phi$ refers to the Bardeen gauge-invariant gravitational potential \citep{Bardeen1980}, which its power spectrum $P_{\Phi}(k)=A/k^{4-n_{s}}$ is normalized to $A^2$. There also exist sub-leading order terms in the above expression that lead to extra non-separable shapes, but these are expected to be much smaller without special fine-tuning. The DBI models of inflation predict the non-linearity parameter as \citep{Alishahiha2004, Chen2007}
\begin{equation}
\label{f_{NL}^{DBI}}
f_{{\rm NL}}^{{\rm DBI}}=-\frac{35}{108}\left(\frac{1}{c_{s}^{2}}-1\right).
\end{equation}
The Planck Collaboration \citep{Planck2015non-Gaussianity} has constrained the non-separable shape given by Equation (\ref{B_{Phi}}) and found $f_{{\rm NL}}^{{\rm DBI}}=2.6\pm61.6$ from temperature data and also $f_{{\rm NL}}^{{\rm DBI}}=15.6\pm37.3$ from both temperature and polarization data at 68\% CL. Apart from this, they have applied their 95\% CL constraints on Equation (\ref{f_{NL}^{DBI}}) and reported the following lower bounds on the sound speed of the DBI inflation as
\begin{align}
 \label{c_s,95,T-{only}}
 c_{s}\geq0.069 \qquad &(95\%\,\mathrm{CL},\,T\mathrm{-only}),
 \\
 \label{c_s,68,T+E}
 c_{s}\geq0.087 \qquad &(95\%\,\mathrm{CL},\,T+E),
\end{align}
where we will consider the lower bound (\ref{c_s,68,T+E}) in the next sections. In what follows, we investigate the power-law, intermediate, and logamediate inflations driven by the DBI scalar field with constant speed of sound and examine the viability of these models in light of the Planck 2015 observational results.

\section{Power-law DBI inflation}
\label{section:power-law}

The first case that we study is the power-law scale factor
\begin{equation}
\label{a,power-law}
a(t) = a_0 t^q,
\end{equation}
where $q>1$ is a constant parameter. This scale factor has already been investigated in the framework of DBI inflation with constant sound speed in \citet{Tsujikawa2013}, where the authors have derived the warp factor and inflationary potential corresponding to this model. They also have presented the bound on the sound speed of the model deduced from the Planck 2013 constraints on the primordial non-Gaussianity. Here, we proceed to examine this model in greater detail, and in particular we are interested in displaying the explicit prediction of this model in the $r-n_s$ plane, which is a crucial test to check the admissibility of each inflationary model in light of the observational data. Also, we specify the parameter space of the power-law DBI inflation for which our model is compatible with the 68\% or 95\% CL regions of the Planck 2015 TT,TE,EE+lowP data.

Considering scale factor (\ref{a,power-law}) and using Equation (\ref{{dot}{{phi}},H'}) for the constant $\gamma$, one can find
\begin{equation}
\label{{dot}{{phi}},power-law}
\dot{\phi}=\frac{\sqrt{2c_{s}q}}{t}.
\end{equation}
The above differential equation can be easily solved to give
\begin{equation}
\label{{phi},t,power-law}
\phi=\sqrt{2c_{s}q}\,\ln t,
\end{equation}
where the integration constant has been put to zero without loss of generality. With the help of Equations (\ref{a,power-law}) and (\ref{{phi},t,power-law}), it is a trivial task to show that the Hubble parameter $H$ is expressed in terms of the DBI scalar field $\phi$ as
\begin{equation}
\label{H,{phi},power-law}
H=q\,\exp{\left(-\frac{\phi}{\sqrt{2c_{s}q}}\right)}.
\end{equation}
For the constant $c_s$, using Equations (\ref{c_s}), (\ref{{dot}{{phi}},power-law}), and (\ref{{phi},t,power-law}) and $X=\dot{\phi}^2/2$, the warp factor acquires the form
\begin{equation}
\label{f,{phi},power-law}
f(\phi)=\frac{\left(1-c_{s}^{2}\right)}{2c_{s}q}\,e^{\sqrt{\frac{2}{c_{s}q}}\,\phi}.
\end{equation}
In the slow-roll approximation, we can use Equation (\ref{H,{phi},power-law}) in the Friedmann equation (\ref{H,V}), and find the inflationary potential as
\begin{equation}
\label{V,phi,power-law}
V(\phi)=3q^{2}e^{-\sqrt{\frac{2}{c_{s}q}}\,\phi}.
\end{equation}
This results reflects the fact that in our DBI framework with constant sound speed, the power-law inflation arises from an exponential potential, and this is in agreement with the result found in \citet{Tsujikawa2013}. Also, in the case of $c_s=1$, the above equation indicates that the power-law inflation is driven by the exponential potential $V(\phi)\propto e^{-\sqrt{\frac{2}{q}}\phi}$, which is the well-known result that we expect in the standard canonical inflationary setting \citep{Lucchin1985, Halliwell1987, Yokoyama1988, Liddle1989}.

Considering the power-law scale factor (\ref{a,power-law}), the first Hubble slow-roll parameter (\ref{{varepsilon}_1}) becomes
\begin{equation}
\label{{varepsilon}_1,power-law}
\varepsilon_1 = \frac{1}{q},
\end{equation}
whereas the other Hubble slow-roll parameters vanish. From the above equation it is evident that $\varepsilon_1$ is constant during the inflationary era, and hence it cannot reach unity at the end of inflation, which means that in this model, inflation cannot terminate by slow-roll violation \citep[i.e. $\varepsilon_1=1$;][]{Martin2014, Rezazadeh2015, Rezazadeh2016, Rezazadeh2017, Rezazadeh2017-2, Zhang2014}.

Substituting Equation (\ref{{varepsilon}_1,power-law}) into (\ref{{mathcal}{P}_s}), the scalar power spectrum is obtained as
\begin{equation}
\label{{mathcal}{P}_s,t,power-law}
{\cal P}_{s}=\frac{q^{3}}{8\pi^{2}c_{s}t^{2}}.
\end{equation}

For the power-law inflation (\ref{a,power-law}), solving the differential equation (\ref{dN}) gives $t$ in terms of the $e$-fold number $N$ as
\begin{equation}
\label{t,N,C,power-law}
t=Ce^{-\frac{N}{q}},
\end{equation}
where $C$ is the constant of integration that can be determined by applying the condition $N_e=0$ at the end time of inflation $t=t_e$, which is a direct consequence of definition (\ref{N}) for the number of $e$-foldings. Thus, we set $C=t_e$ and consequently
\begin{equation}
\label{t,N,power-law}
t=t_{e}e^{-\frac{N}{q}}.
\end{equation}
By inserting Equation (\ref{t,N,power-law}) into (\ref{{mathcal}{P}_s,t,power-law}), the power spectrum of scalar density perturbations is given in terms of the $e$-fold number $N$ as
\begin{equation}
\label{{mathcal}{P}_s,N,power-law}
{\cal P}_{s}=\frac{q^{3}}{8\pi^{2}c_{s}t_{e}^{2}}~e^{\frac{2N}{q}}.
\end{equation}
Fixing the amplitude of the scalar power spectrum in the above equation at the horizon crossing, one can determine the end time of inflation $t_e$ in terms of the other parameters of the model. Using Equation (\ref{{varepsilon}_1,power-law}) in (\ref{n_s}) and keeping in mind that $\varepsilon_2=\varepsilon_{s1}=0$, the scalar spectral index is obtained as
\begin{equation}
\label{n_s,power-law}
n_{s}=1-\frac{2}{q}.
\end{equation}
As we see, the result (\ref{n_s,power-law}) does not depend on the $e$-fold number $N$, and hence it leads to a vanishing running of the scalar spectral index,
\begin{equation}
 \label{dn_s/d{ln}k,power-law}
 \frac{dn_{s}}{d\ln k}=0,
\end{equation}
satisfying the 95\% CL constraint deduced from the Planck 2015 TT,TE,EE+lowP data \citep{Planck2015}.

Using Equations (\ref{r}) and (\ref{{varepsilon}_1,power-law}), the tensor-to-scalar ratio will be
\begin{equation}
\label{r,power-law}
r=\frac{16c_{s}}{q}.
\end{equation}
We see that $r$ like $n_s$, is independent of $N$, and therefore we can combine Equations (\ref{n_s,power-law}) and (\ref{r,power-law}) to reach
\begin{equation}
\label{r,n_s,power-law}
r=8c_{s}\left(1-n_{s}\right),
\end{equation}
implying a linear relation between $r$ and $n_s$. The existence of a linear relation between these two observables for the power-law inflation also appears in other inflationary scenarios, such as the standard canonical inflation \citep{Tsujikawa2013}, the tachyon inflation \citep{Rezazadeh2017-2}, and the Brans-Dicke inflation \citep{Tahmasebzadeh2016}. Note that we can only combine the equations for $n_s$ and $r$ to eliminate the constant parameters, and this is not allowed to do this for the dynamical variables like the cosmic time $t$, the scalar field $\phi$, or the $e$-fold number $N$ \citep{Rezazadeh2017-2}. Otherwise, one may incorrectly avoid the essential requirement that the inflationary observables must be evaluated at the time of horizon exit \citep{Rezazadeh2017-2}.

Now, we use Equations (\ref{n_s,power-law}) and (\ref{r,power-law}) to plot the $r-n_s$ diagram of our DBI power-law inflation model as demonstrated in Figure \ref{figure:r,n_s,power-law}. In this figure, the prediction of the model is shown by the black lines for different values of sound speed $c_s$ while the parameter $q$ is taken to be varying in the range $q>1$. It should be noted that the selected values for $c_s$ satisfy the sound speed bound $c_{s}\geq0.087$ obtained in the previous section due to the 95\% CL constraint of the Planck 2015 data on the non-Gaussianity parameter $f_{{\rm NL}}^{{\rm DBI}}$. Also, in this figure, the marginalized joint 68\% and 95\% CL regions of the Planck data \citep{Planck2015} have been specified. From the figure, it is clear that for the case of $c_s=1$ corresponding to the power-law inflation in the standard canonical scalar field framework, the result of model is completely ruled out by the the Planck 2015 TT,TE,EE+lowP data \citep{Planck2015}. But taking $0.083 \leq c_s \lesssim 0.29$ and $0.29 \lesssim c_s \lesssim 0.63$, the predictions of power-law inflation in DBI non-canonical scalar field model are compatible with the joint 68\% and 95\% CL regions of the the Planck 2015 TT,TE,EE+lowP data \citep{Planck2015}, respectively.

\begin{figure*}
\begin{center}
\scalebox{1}[1]{\includegraphics{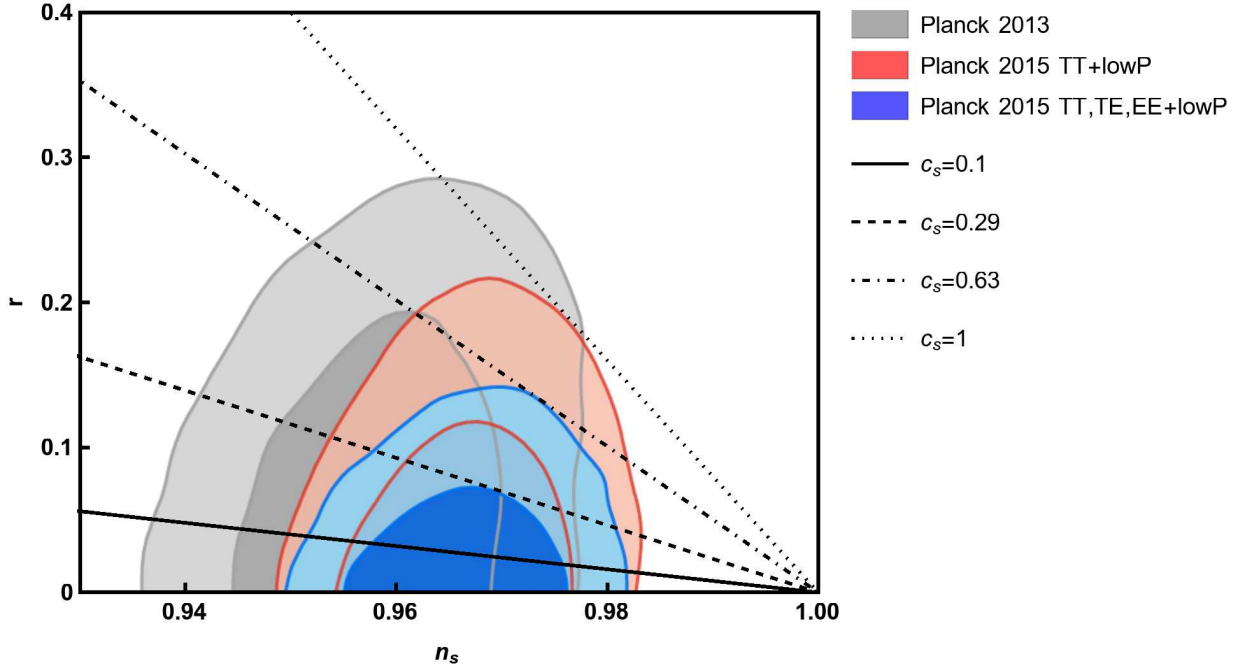}}
\caption{The $r-n_s$ diagram for the power-law inflation (\ref{a,power-law}) in the DBI scenario (\ref{{mathcal}{L}}) for different values of constant sound speed $c_s$ with varying $q$ in the range $q>1$. The marginalized joint 68\% and 95\% CL regions of Planck 2013, Planck 2015 TT+lowP, and the Planck 2015 TT,TE,EE+lowP data \citep{Planck2015} are specified by gray, red, and blue, respectively.}
\label{figure:r,n_s,power-law}
\end{center}
\end{figure*}

In Table \ref{table:power-law}, we further have determined the intervals of $q$ that our model with some specified values of $c_s$ can be in agreement with the Planck 2015 TT,TE,EE+lowP data. In order to specify the allowed ranges of the parameter $q$ in Table \ref{table:power-law}, we locate the prediction of the model in the $r-n_s$ plane for different values of $q$, and find the values of $q$ for which the result of the model lies on the boundaries of the 68\% or 95\% CL regions of the CMB data.

\begin{table}
  \centering
  \caption{The Ranges of Parameter $q$ for which the $r-n_s$ Diagram of the DBI Power-law Inflation with Different Values of the Constant Sound Speed $c_s$ Is Compatible with the 68\% or 95\% CL Regions of the Planck 2015 TT,TE,EE+lowP Data \citep{Planck2015}}
\scalebox{1}{
\begin{tabular}{ccc}
  \hline
  \hline
  $c_s$ & $\qquad$ $q$ (68\% CL) $\qquad$ & $q$ (95\% CL) \\
  \hline
  $0.1$ & $[47, 83]$ & $[41, 111]$ \\
  $0.2$ & $[53, 78]$ & $[44, 109]$ \\
  $0.3$ & --- & $[48, 104]$ \\
  $0.4$ & --- & $[52, 101]$ \\
  $0.5$ & --- & $[58, 97]$ \\
  $0.6$ & --- & $[67, 83]$ \\
  $c_s \gtrsim 0.63$ & --- & --- \\
  \hline
\end{tabular}
}
\label{table:power-law}
\end{table}

It is useful to determine the parameter space of $q$ and $c_s$ for consistency of our model with the Planck 2015 constraints. This parameter space is represented in Figure \ref{figure:q,c_s,power-law}, where the darker and lighter blue areas point to the regions for which the $r-n_s$ result of the model verifies, respectively, the 68\% and 95\% CL constraints of the Planck 2015 TT,TE,EE+lowP data \citep{Planck2015}. To draw Figure \ref{figure:q,c_s,power-law}, we have provided a computational code to calculate $n_s$ and $r$ by using of Equations (\ref{n_s,power-law}) and (\ref{r,power-law}) for given values of $q$ and $c_s$, and then we check weather or not the prediction of the model can lie inside the 68\% or 95\% CL regions of the observational data. In this way, we can separate the values of $q$ and $c_s$ for which the predictions of the model are consistent with the observations, and these values can be projected in a counter-plot as shown in Figure \ref{figure:q,c_s,power-law}.

\begin{figure}
\begin{center}
\scalebox{0.8}[0.8]{\includegraphics{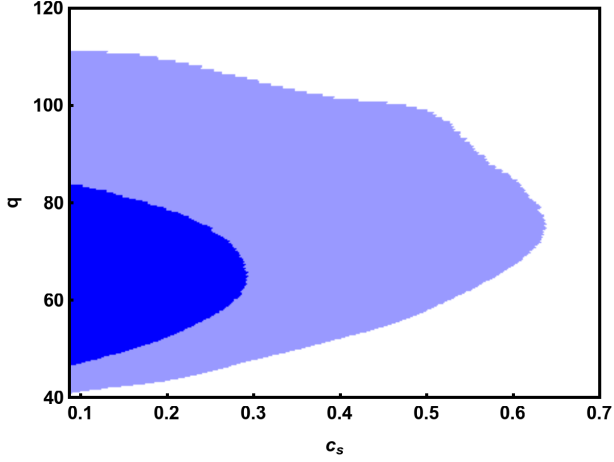}}
\caption{Parameter space of $q$ and $c_s$, for which the power-law inflation (\ref{a,power-law}) in the DBI framework (\ref{{mathcal}{L}}) is consistent with the Planck 2015 results. The darker and lighter blue regions specify the parameter space for which the $r-n_s$ prediction of model fulfills, respectively, the 68\% and 95\% CL constraints of the Planck 2015 TT,TE,EE+lowP data \citep{Planck2015}.}
\label{figure:q,c_s,power-law}
\end{center}
\end{figure}

In order to compare the results of our model with the observational data more quantitatively, we evaluate the inflationary observables, including $n_s$, $r$, and $f_{{\rm NL}}^{{\rm DBI}}$, with the help of Equations (\ref{n_s,power-law}), (\ref{r,power-law}), and (\ref{f_{NL}^{DBI}}), respectively. These observables are estimated for different values of $c_s$ and $q$, and the results are presented in Table \ref{table:power-law,observables}. Using the obtained values for $n_s$ and $r$, we check the consistency of the model in the $r-n_s$ plane in comparison with the Planck 2015 TT,TE,EE+lowP data \citep{Planck2015}, and summarize our results in the last column of Table \ref{table:power-law,observables}. It should be noted that the quantitative results of Table \ref{table:power-law,observables} verify the qualitative findings of Figure \ref{figure:q,c_s,power-law}. The values of $f_{{\rm NL}}^{{\rm DBI}}$ presented in the fifth column of Table \ref{table:power-law,observables} are compatible with the lower bound $c_{s}\geq0.087$ imposed by the 95\% CL constraint of the Planck 2015 T+E data \citep{Planck2015non-Gaussianity} on the non-Gaussianity parameter.

\begin{table*}
  \centering
  \caption{Estimated values of inflationary observables for the DBI power-law inflation with some typical values of $c_s$ and $q$.}
\scalebox{1}{
\begin{tabular}{cccccc}
  \hline
  \hline
  $\qquad$ $c_s$ $\qquad$ & $\qquad$ $q$ $\qquad$ & $\qquad$ $n_s$ $\qquad$ & $\qquad$ $r$ $\qquad$ & $\qquad$ $f_{{\rm NL}}^{{\rm DBI}}$ $\qquad$ & $r-n_s$ Consistency \\
  \hline
  $0.1$ & $65$ & $0.9692$ & $0.0246$ & $-32.083$ & 68\% CL \\
  $0.2$ & $65.5$ & $0.9695$ & $0.0489$ & $-7.778$ & 68\% CL \\
  $0.3$ & $76$ & $0.9737$ & $0.0632$ & $-3.277$ & 95\% CL \\
  $0.4$ & $76.5$ & $0.9739$ & $0.0837$ & $-1.701$ & 95\% CL \\
  $0.5$ & $77.5$ & $0.9742$ & $0.1032$ & $-0.972$ & 95\% CL \\
  $0.6$ & $75$ & $0.9733$ & $0.1280$ & $-0.576$ & 95\% CL \\
  $0.7$ & $80$ & $0.9750$ & $0.1400$ & $-0.337$ & --- \\
  \hline
\end{tabular}
}
\label{table:power-law,observables}
\end{table*}

\section{Intermediate DBI inflation}
\label{section:intermediate}

The next model that we investigate in the DBI scenario is the intermediate scale factor \citep{Barrow1990, Barrow1993, Barrow2006, Barrow2007}
\begin{equation}
\label{a,intermediate}
a(t)=a_{0}\exp\left(At^{\lambda}\right),
\end{equation}
where $a_0>0$, $A>0$, and $0<\lambda<1$ are constant parameters. Note that the intermediate inflation driven by a DBI scalar field has already been investigated in \citet{Nazavari2016}, but here we follow a quite different approach. In \citet{Nazavari2016}, the authors have chosen the AdS warp factor $f(\phi)=\lambda/\phi^{4}$ in their investigation, while our work is based on the assumption of constant sound speed, which enables us to determine the warp factor as well as the inflationary potential associated with the intermediate scale factor (\ref{a,intermediate}).

With the intermediate scale factor (\ref{a,intermediate}), the Hubble parameter becomes
\begin{equation}
\label{H,t,intermediate}
H=\frac{A\lambda}{t^{1-\lambda}}.
\end{equation}
For the constant sound speed, inserting Equation (\ref{H,t,intermediate}) into (\ref{{dot}{{phi}},H'}) yields
\begin{equation}
\label{{dot}{{phi}},intermediate}
\dot{\phi}=\sqrt{\frac{2c_{s}A\lambda\left(1-\lambda\right)}{t^{2-\lambda}}}.
\end{equation}
Integrating the above equation gives
\begin{equation}
\label{{phi},t,intermediate}
\phi=2\sqrt{\frac{2c_{s}A\left(1-\lambda\right)}{\lambda}}~t^{\lambda/2}.
\end{equation}
By replacing $t$ from Equation (\ref{{phi},t,intermediate}) into (\ref{H,t,intermediate}), we reach
\begin{equation}
\label{H,{phi},intermediate}
H=\lambda^{2}\left(\frac{A}{\lambda}\right)^{\frac{1}{\lambda}}\Big[8c_{s}(1-\lambda)\Big]^{\frac{1-\lambda}{\lambda}}\phi^{-\frac{2(1-\lambda)}{\lambda}}.
\end{equation}
With the help of Equations (\ref{c_s}), (\ref{{dot}{{phi}},intermediate}), and (\ref{{phi},t,intermediate}), and also using $X=\dot{\phi}^2/2$, the background warp factor can be found as
\begin{equation}
\label{f,{phi},intermediate}
f(\phi)=4^{-\frac{\lambda-3}{\lambda}}\left(1-c_{s}^{2}\right)\left[\frac{\lambda^{1-\lambda}}{c_{s}A\left(1-\lambda\right)}\right]^{\frac{2}{\lambda}}\phi^{\frac{2\left(2-\lambda\right)}{\lambda}}.
\end{equation}
In addition, inserting Equation (\ref{H,{phi},intermediate}) into the first Friedmann equation (\ref{H,V}), the brane potential for this model in the slow-roll approximation reads
\begin{eqnarray}
\label{V,{phi},intermediate}
V(\phi)=3\lambda^{4}\left(\frac{A}{\lambda}\right)^{\frac{2}{\lambda}}\Big[8c_{s}(1-\lambda)\Big]^{\frac{2}{\lambda}}\phi^{-\frac{4(1-\lambda)}{\lambda}}.
\end{eqnarray}
This equation implies that in our DBI setting, like the canonical framework \citep{Barrow1990, Barrow1993, Barrow2006, Barrow2007}, the intermediate inflation (\ref{a,intermediate}) stems from the inverse power-law potential $V(\phi)\propto\phi^{-4(1-\lambda)/\lambda}$.

Considering the Hubble parameter (\ref{H,t,intermediate}), the Hubble slow-roll parameters (\ref{{varepsilon}_1}) and (\ref{{varepsilon}_{i+1}}) read
\begin{eqnarray}
\label{{varepsilon}_1,t,intermediate}
\varepsilon_{1} &=& \frac{\left(1-\lambda\right)}{A\lambda}\;t^{-\lambda},
\\
\label{{varepsilon}_{i+1},t,intermediate}
\varepsilon_{i+1} &=& -\frac{1}{A}\;t^{-\lambda},\qquad(i\geq1).
\end{eqnarray}
Using Equations (\ref{H,t,intermediate}) and (\ref{{varepsilon}_1,t,intermediate}) in (\ref{{mathcal}{P}_s}), we obtain the power spectrum of primordial density perturbations as
\begin{equation}
\label{{mathcal}{P}_s,t,intermediate}
{\cal P}_{s}=\frac{\left(A\lambda\right)^{3}}{8\pi^{2}c_{s}\left(1-\lambda\right)}\,t^{3\lambda-2}.
\end{equation}
By integrating the differential equation $dN=-Hdt$ in which $H$ is now given by Equation (\ref{H,t,intermediate}), one can express the cosmic time $t$ in terms of the $e$-folds number $N$ for the intermediate inflation as
\begin{equation}
\label{t,N,intermediate}
t=\left(t_{e}^{\lambda}-\frac{N}{A}\right)^{\frac{1}{\lambda}},
\end{equation}
where $t_e$ refers to the end time of inflation. Now, by replacing $t$ from Equation (\ref{t,N,intermediate}) into (\ref{{mathcal}{P}_s,t,intermediate}), the power spectrum of scalar perturbations acquires the form
\begin{equation}
\label{{mathcal}{P}_s,N,intermediate}
{\cal P}_{s}=\frac{\left(A\lambda\right)^{3}}{8\pi^{2}c_{s}(1-\lambda)}\left(t_{e}^{\lambda}-\frac{N}{A}\right)^{\frac{3\lambda-2}{\lambda}}.
\end{equation}
Also, using Equations (\ref{{varepsilon}_1,t,intermediate}), (\ref{{varepsilon}_{i+1},t,intermediate}), and (\ref{t,N,intermediate}) in (\ref{n_s}), the scalar spectral index reads
\begin{equation}
\label{n_s,N,intermediate}
n_{s}=1-\frac{2-3\lambda}{A\lambda\left(t_{e}^{\lambda}-\frac{N}{A}\right)},
\end{equation}
where we have set $\varepsilon_{s1}=0$, because the sound speed is constant in our model. It should be noted that the parameter $\lambda$ should be in the range of $0<\lambda<2/3$, because for $2/3<\lambda<1$ we obtain a blue-tilted spectrum ($n_s>1$) that is completely ruled out by the Planck 2015 data \citep{Planck2015}. In addition, for $\lambda=2/3$, we obtain the scale-invariant Harrison-Zel'dovich spectrum ($n_s=0$) which is also at odds with the Planck 2015 observations \citep{Planck2015}.

By applying Equations (\ref{{varepsilon}_1,t,intermediate}), (\ref{{varepsilon}_{i+1},t,intermediate}), and (\ref{t,N,intermediate}) in (\ref{dn_s/d{ln}k}), and keeping in mind that $\varepsilon_{s1}=\varepsilon_{s2}=0$, we can further calculate the running of the scalar spectral index for our model as
\begin{equation}
\label{dn_s/d{ln}k,N,intermediate}
\frac{dn_{s}}{d\ln k}=\frac{2-3\lambda}{A^{2}\lambda\left(t_{e}^{\lambda}-\frac{N}{A}\right)}.
\end{equation}
In order to find the relation of the tensor-to-scalar ratio in our model, we use Equations (\ref{{varepsilon}_1,t,intermediate}) and (\ref{t,N,intermediate}) in (\ref{r}), and we will have
\begin{equation}
\label{r,N,intermediate}
r = \frac{16c_s(1 - \lambda)}{A\lambda\left(t_e^\lambda-\frac{N}{A}\right)}.
\end{equation}

Now, we are interested in determining the parameter $t_e$ in terms of the other parameters of the model. For this purpose, we evaluate the scalar power spectrum (\ref{{mathcal}{P}_s,N,intermediate}) at the epoch of horizon crossing with the $e$-fold number $N_*$. Solving the resulting equation for $t_e$ yields
\begin{equation}
 \label{t_e,intermediate}
 t_{e}=\left[\left(\frac{8\pi^{2}c_{s}(1-\lambda)}{\left(A\lambda\right)^{3}}\mathcal{P}_{s*}\right)^{\frac{\lambda}{3\lambda-2}}+\frac{N_{*}}{A}\right]^{\frac{1}{\lambda}},
\end{equation}
where $\ensuremath{\mathcal{P}_{s*}\equiv\left.\mathcal{P}_{s}\right|_{N=N_{*}}}$ is a constant number given by $\ln\left[10^{10}{\cal P}_{s*}\right]=3.094\pm0.034$ according to the 68\% CL constraint from the Planck 2015 TT,TE,EE+lowP data \citep{Planck2015}. Substituting Equation (\ref{t_e,intermediate}) into Equations (\ref{n_s,N,intermediate}), (\ref{dn_s/d{ln}k,N,intermediate}), and (\ref{r,N,intermediate}), we find, respectively,
\begin{eqnarray}
 \label{n_s,intermediate}
 n_{s}&=& 1-\left(2-3\lambda\right)\left(\frac{8\pi^{2}c_{s}(1-\lambda)}{\left(A\lambda\right)^{2/\lambda}}{\cal P}_{s*}\right)^{\frac{\lambda}{2-3\lambda}},
 \\
 \label{dn_s/d{ln}k,intermediate}
 \frac{dn_{s}}{d\ln k} &=& \frac{2-3\lambda}{A^{2}\lambda}\left(\frac{8\pi^{2}c_{s}(1-\lambda)}{\left(A\lambda\right)^{3}}{\cal P}_{s*}\right)^{\frac{2\lambda}{2-3\lambda}},
 \\
 \label{r,intermediate}
 r&=& 16c_{s}(1-\lambda)\left(\frac{8\pi^{2}c_{s}(1-\lambda)}{\left(A\lambda\right)^{2/\lambda}}{\cal P}_{s*}\right)^{\frac{\lambda}{2-3\lambda}}.
\end{eqnarray}
The above observables are completely independent of the dynamical quantities such as the cosmic time $t$, the inflaton field $\phi$, and the $e$-fold number $N$. As a result, we can easily combine Equations (\ref{n_s,intermediate}) and (\ref{r,intermediate}) to find
\begin{equation}
 \label{r,n_s,intermediate}
 r=\frac{16c_{s}(1-\lambda)}{2-3\lambda}\left(1-n_{s}\right),
\end{equation}
which demonstrates a linear relation between $r$ and $n_s$ in our DBI intermediate model. With $c_s=1$, this is in agreement with the results found in \citet{Barrow2006} and \citet{Barrow2007} for the intermediate inflation in the standard canonical setup. It is interesting that for $\lambda\rightarrow0$, the relation (\ref{r,n_s,intermediate}) reduces to the linear relation (\ref{r,n_s,power-law}) obtained in the previous section for the DBI power-law inflation.

We use Equations (\ref{n_s,intermediate}) and (\ref{r,intermediate}) to plot the $r-n_s$ diagram for the intermediate inflation (\ref{a,intermediate}) driven by the DBI scalar field (\ref{{mathcal}{L}}). This diagram is illustrated in Figure \ref{figure:r,n_s,intermediate} for some typical values of $c_s$ and $\lambda$ with varying $A$ in the range of $A>0$. From Figure \ref{figure:r,n_s,intermediate}, it is clear that for the case $c_s=1$ corresponding to the intermediate inflation in the standard canonical setting, the result of model lies entirely outside the observational regions allowed by the Planck 2015 data \citep{Planck2015}. Also, it should be noted that the $r-n_s$ diagram for $c_s=1$ confirms the one presented in \citet{del-Campo2014} for the intermediate inflation in the standard canonical framework. However, as we infer from the figure, if we choose the sound speed $c_s$ sufficiently smaller than the light speed ($c=1$), then our model can be compatible with the observational data. For instance, as shown in Figure \ref{figure:r,n_s,intermediate}, the prediction of model with $c_s=0.1$ can lie inside the 68\% CL region of the Planck 2015 TT,TE,EE+lowP data \citep{Planck2015}. More precisely, we deduce from the figure that in the case of $c_s=0.1$, our intermediate model is favored according to the 68\% CL (95\% CL) constraints of the aforementioned data, if we take $\lambda \lesssim 0.565$ ($\lambda \lesssim 0.627$).

\begin{figure*}
\begin{center}
\scalebox{1}[1]{\includegraphics{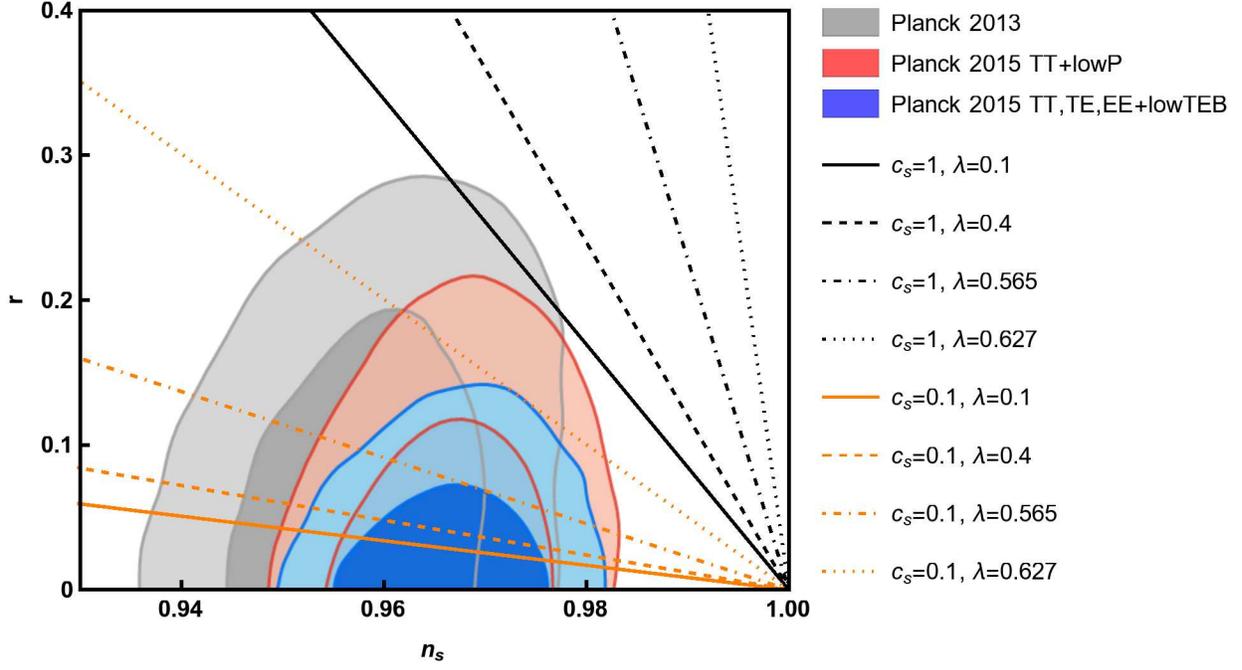}}
\caption{Same as Figure \ref{figure:r,n_s,power-law}, but for the intermediate inflation (\ref{a,intermediate}) in the DBI setting (\ref{{mathcal}{L}}) for some typical values of $c_s$ and $\lambda$ with varying $A$ in the range of $A>0$. }
\label{figure:r,n_s,intermediate}
\end{center}
\end{figure*}

\begin{table*}
  \centering
  \caption{The ranges of Parameter $A$, for which the $r-n_s$ diagram of the DBI intermediate inflation with $c_s=0.1$ and some typical values of $\lambda$ is compatible with the 68\% or 95\% CL regions of the Planck 2015 TT,TE,EE+lowP data \citep{Planck2015}. Furthermore, the estimated values for the running of the scalar spectral index $dn_s/d\ln k$ are presented in the table that satisfy the 95\% CL constraint of the Planck 2015 TT,TE,EE+lowP data \citep{Planck2015}.}
\scalebox{0.8}{
\begin{tabular}{ccccc}
  \hline
  \hline
  $\quad$  $\lambda$ $\quad$ & $\quad\quad\quad$ $A$ (68\% CL) $\quad\quad\quad$ & $\quad\quad\quad$ $\frac{dn_s}{d\ln k}$ $\quad\quad\quad$ & $\quad\quad\quad$ $A$ (95\% CL) $\quad\quad\quad$ & $\quad\quad\quad$ $\frac{dn_s}{d\ln k}$ $\quad\quad\quad$ \\
  \hline
  $0.1$ & $[94.5,150]$ & $[3.507 \times 10^{-5},1.040 \times 10^{-4}]$ & $[84,195]$ & $[1.892 \times 10^{-5},1.372 \times 10^{-4}]$\\
  $0.2$ & $[9.5,14]$ & $[8.415 \times 10^{-5},2.548 \times 10^{-4}]$ & $[8.7,17]$ & $[4.832 \times 10^{-5},3.276 \times 10^{-4}]$ \\
  $0.3$ &$[1.32,1.75]$ & $[1.658 \times 10^{-4},4.622 \times 10^{-4}]$ & $[1.215,2.06]$ & $[9.162 \times 10^{-5},6.248 \times 10^{-4}]$ \\
  $0.4$ & $[0.209,0.254]$ & $[3.090 \times 10^{-4},8.193 \times 10^{-4}]$ & $[0.196,0.287]$ & $1.678\times 10^{-4},1.130 \times 10^{-3}]$ \\
  $0.5$ & $[0.0368,0.0405]$ & $[6.712 \times 10^{-4},1.445 \times 10^{-3}]$ & $[0.0351,0.0439]$ & $[3.522 \times 10^{-4},2.109 \times 10^{-3}]$ \\
  $0.6$ & --- & --- & $[0.00702, 0.00750]$ & $[1.187 \times 10^{-3}, 4.456 \times 10^{-3}]$ \\
  $\lambda \gtrsim 0.627$ & --- & --- & --- & --- \\
  \hline
\end{tabular}
}
\label{table:intermediate}
\end{table*}

\begin{table*}
  \centering
  \caption{Estimated values of inflationary observables for the DBI intermediate inflation with $\lambda=0.1$ and some typical values of $c_s$ and $A$.}
\scalebox{0.8}{
\begin{tabular}{ccccccc}
  \hline
  \hline
  $\qquad$  $c_s$ $\qquad$ & $\qquad$ $A$ $\qquad$ & $\qquad$ $n_s$ $\qquad$ & $\frac{dn_s}{d\ln k}$ & $\qquad$ $r$ $\qquad$ & $\qquad$ $f_{{\rm NL}}^{{\rm DBI}}$ $\qquad$ & $r-n_s$ Consistency \\
  \hline
  $0.1$ & $100$ & $0.9607$ & $9.106 \times 10^{-5}$ & $0.0333$ & $-32.083$ & 68\% CL \\
  $0.2$ & $120$ & $0.9669$ & $6.433 \times 10^{-5}$ & $0.0560$ & $-7.778$ & 68\% CL \\
  $0.3$ & $120$ & $0.9661$ & $6.748 \times 10^{-5}$ & $0.0861$ & $-3.277$ & 95\% CL \\
  $0.4$ & $150$ & $0.9735$ & $4.129 \times 10^{-5}$ & $0.0898$ & $-1.701$ & 95\% CL \\
  $0.5$ & $150$ & $0.9732$ & $4.239 \times 10^{-5}$ & $0.1137$ & $-0.972$ & 95\% CL \\
  $0.6$ & $152$ & $0.9733$ & $4.198 \times 10^{-5}$ & $0.1358$ & $-0.576$ & 95\% CL \\
  $0.7$ & $150$ & $0.9726$ & $4.410 \times 10^{-5}$ & $0.1623$ & $-0.337$ & --- \\
  \hline
\end{tabular}
}
\label{table:intermediate,observables}
\end{table*}

In Table \ref{table:intermediate}, we have tabulated the range of parameter $A$ for which our model with $c_s=0.1$ and with some typical values of $\lambda$ is consistent with the Planck 2015 results. In order to find the allowed intervals of the parameter $A$ reported in Table \ref{table:intermediate}, using Equations (\ref{n_s,intermediate}) and (\ref{r,intermediate}) we determine the values of $A$ for which the results of the model in the $r-n_s$ plane stand on the boundary of the 68\% or 95\% CL regions of the observational data. In addition, in Table \ref{table:intermediate} we have estimated the running of the scalar spectral index $dn_s/d\ln k$ for our model by the use of Equation (\ref{dn_s/d{ln}k,intermediate}). The estimated values for this observable fulfill the 95\% CL constraint of the Planck 2015 TT,TE,EE+lowP data \citep{Planck2015}.

In Figure \ref{figure:A,lambda,intermediate}, we have specified the parameter space of $A$ and $\lambda$ for which the $r-n_s$ result of our model with some specified values of $c_s$ is compatible with the 68\% or 95\% CL regions of the Planck 2015 TT,TE,EE+lowP data \citep{Planck2015}. In order to get this diagram, we have employed a numerical code similar to the one used to provide Figure \ref{figure:q,c_s,power-law}. Here, our code computes $n_s$ and $r$ by using of Equations (\ref{n_s,intermediate}) and (\ref{r,intermediate}), respectively, and then projects the values of $A$ and $\lambda$ which are compatible with the 68\% or 95\% CL regions of the observational data in a two-dimensional counter-plot as shown in Figure \ref{figure:A,lambda,intermediate}.

\begin{figure*}
\begin{minipage}[b]{1\textwidth}
\subfigure{ \includegraphics[width=.48\textwidth]%
{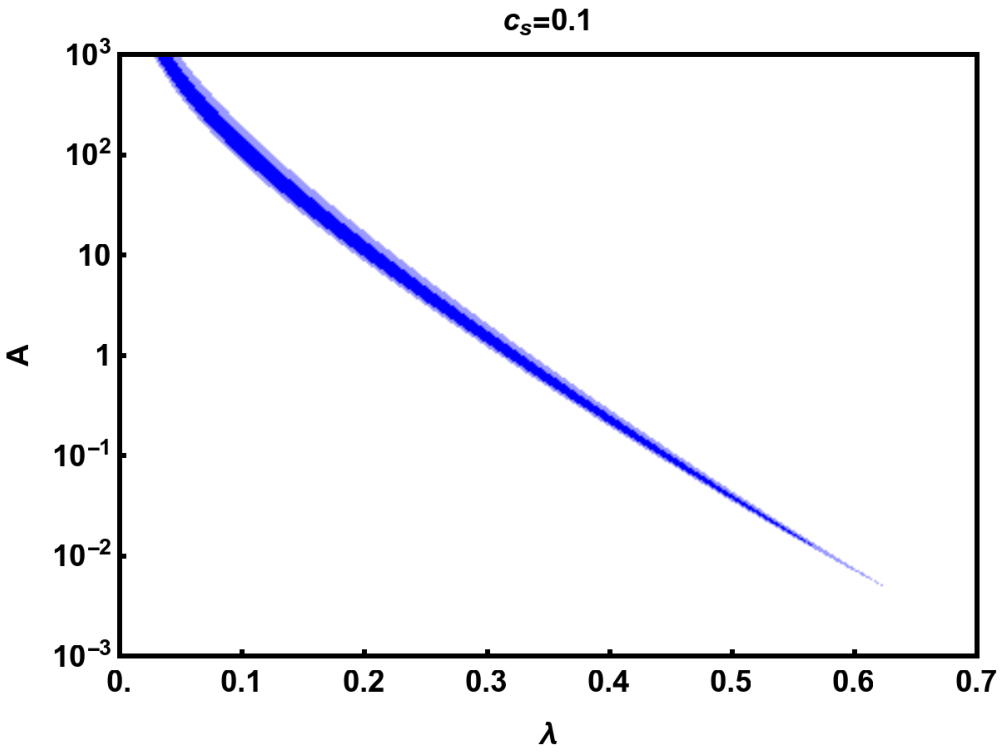}} \hspace{.1cm}
\subfigure{ \includegraphics[width=.48\textwidth]%
{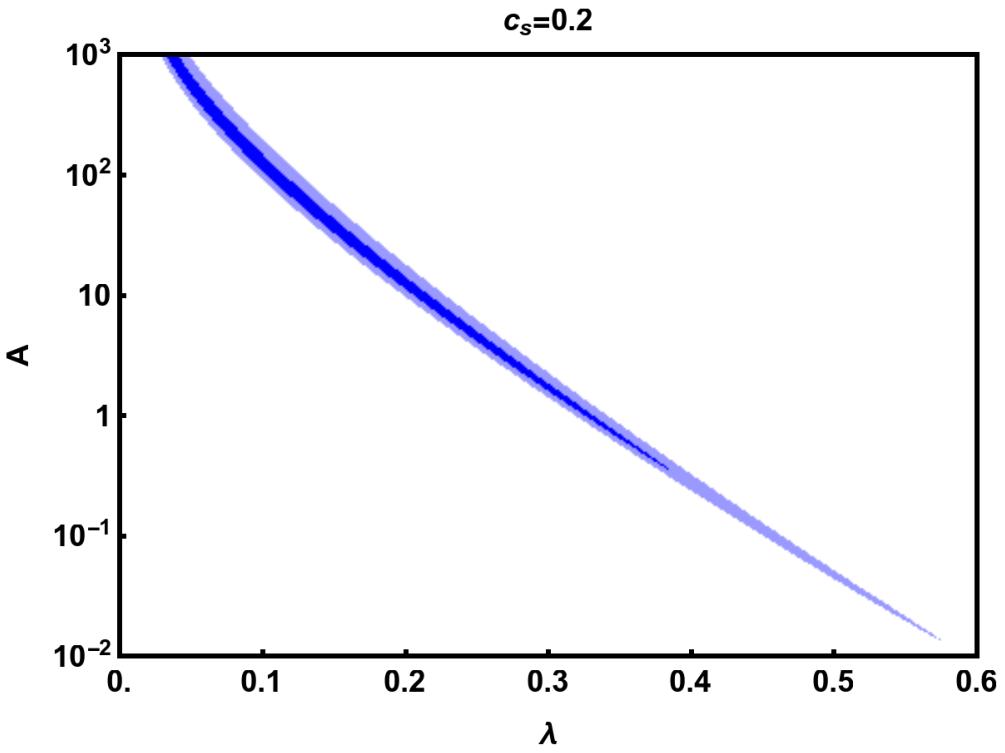}}
\end{minipage}
\begin{minipage}[b]{1\textwidth}
\subfigure{ \includegraphics[width=.48\textwidth]%
{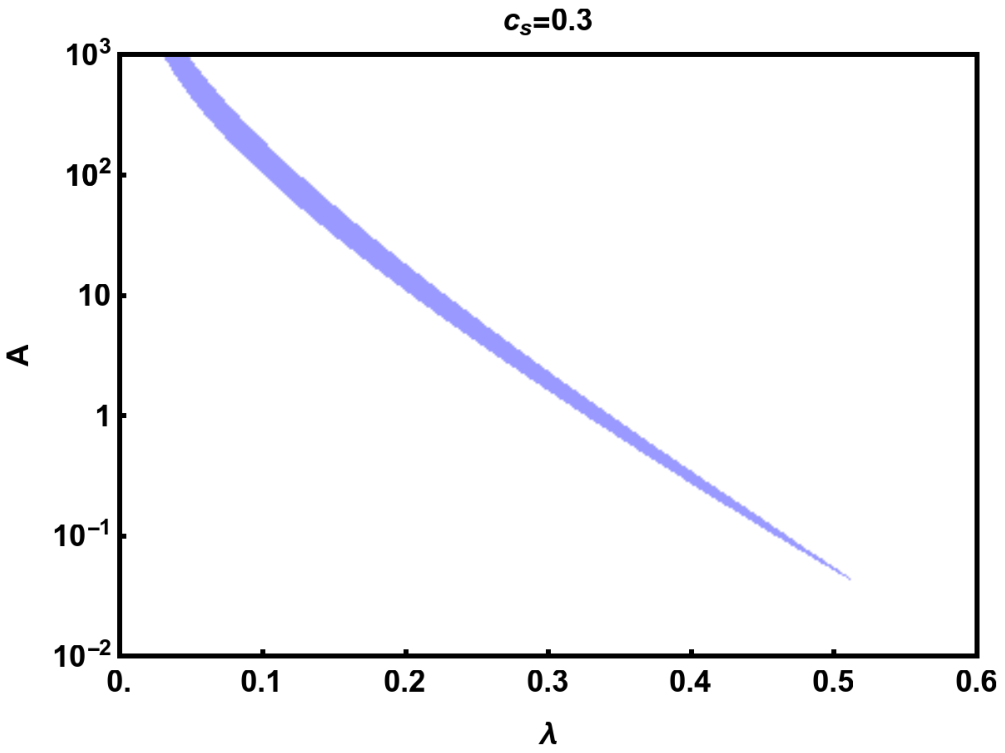}} \hspace{.1cm}
\subfigure{ \includegraphics[width=.48\textwidth]%
{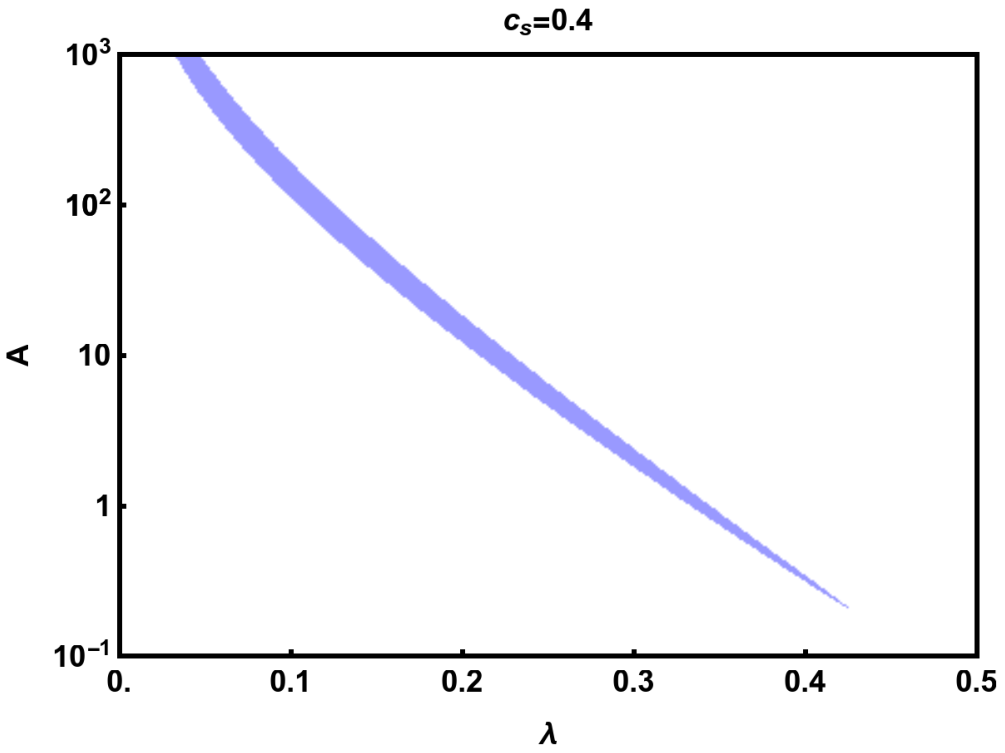}}
\end{minipage}
\begin{minipage}[b]{1\textwidth}
\subfigure{ \includegraphics[width=.48\textwidth]%
{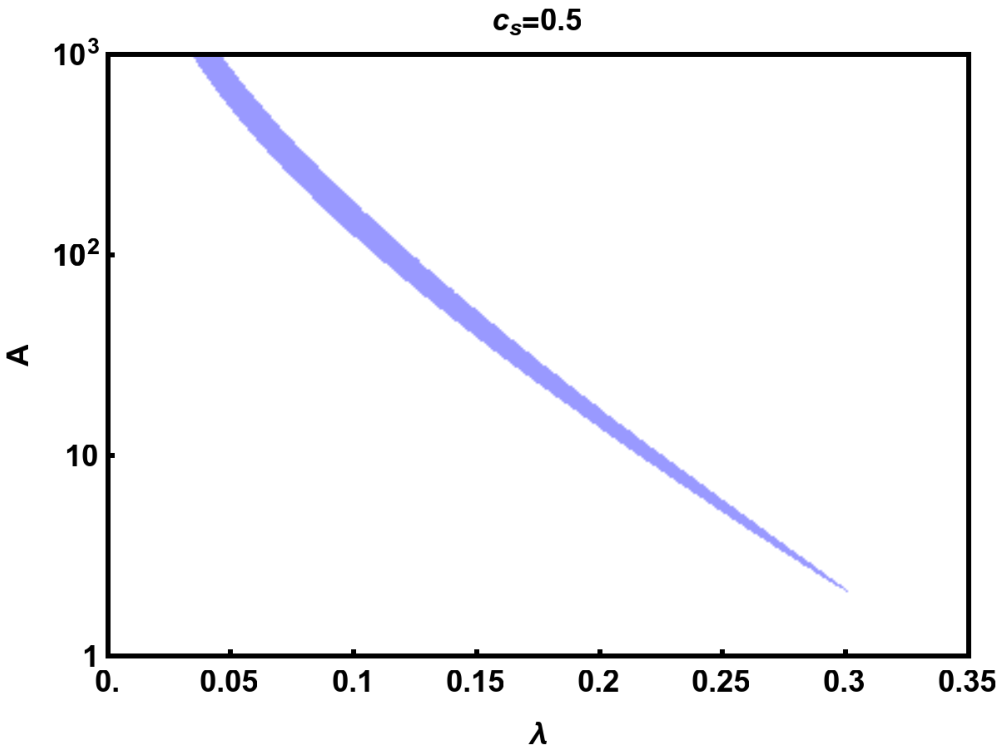}} \hspace{.1cm}
\subfigure{ \includegraphics[width=.48\textwidth]%
{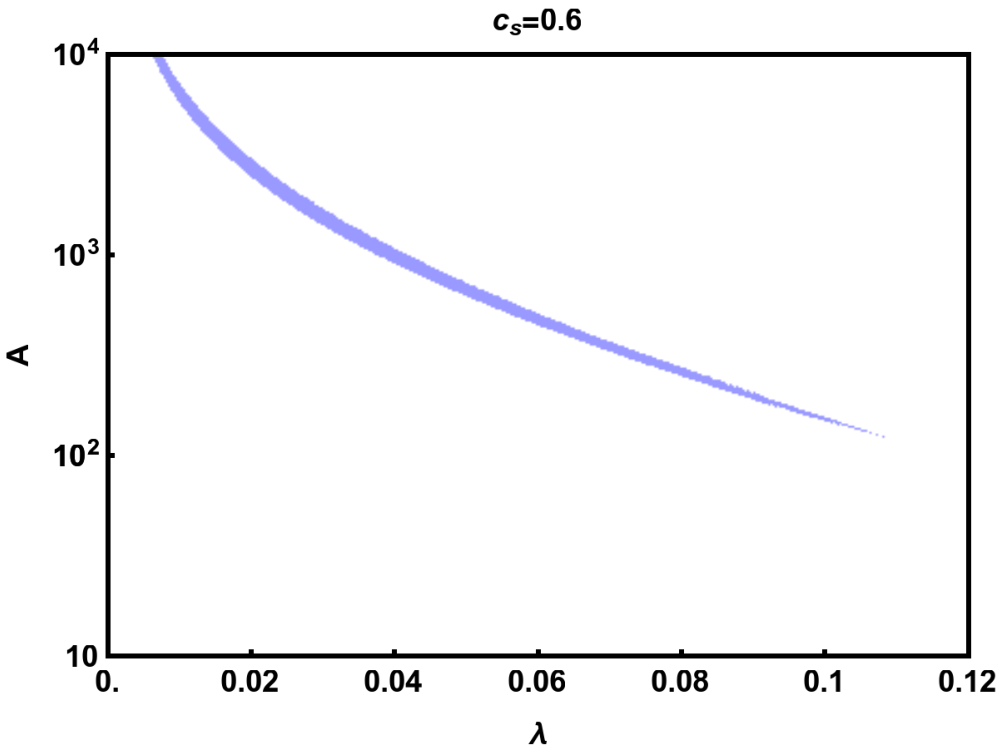}}
\end{minipage}
\caption{Parameter space of $A$ and $\lambda$, for which the intermediate inflation (\ref{a,intermediate}) in the DBI framework (\ref{{mathcal}{L}}) with different values of the constant sound speed $c_s$ is compatible with the Planck 2015 results. The darker and lighter blue regions indicate the parameter space for which the prediction of model in the $r-n_s$ plane is in agreement with the Planck 2015 TT,TE,EE+lowP data \citep{Planck2015} at 68\% and 95\% CL, respectively.}
\label{figure:A,lambda,intermediate}
\end{figure*}

It is useful to compare the predictions of the DBI intermediate model with the CMB data in a more quantitative manner. In Table \ref{table:intermediate,observables}, we have tabulated the inflationary observables for $\lambda =0.1$ and some typical values of $c_s$ and $A$. We have checked the $r-n_s$ consistency of each case and presented our findings in the last column of Table \ref{table:intermediate,observables}, which are in agreement with the results illustrated in Figure \ref{figure:A,lambda,intermediate}.

\section{Logamediate DBI inflation}
\label{section:logamediate}

The last model that we study within the framework of the DBI scalar field is the logamediate inflation characterized by the scale factor \citep{Barrow2007}
\begin{equation}
 \label{a,logamediate}
 a(t)=a_{0}\exp\Big[A\left(\ln t\right)^{\lambda}\Big],
\end{equation}
where $a_0>0$, $A>0$, and $\lambda \geq 1$ are constant parameters. In the case of $\lambda=1$, the logamediate scale factor (\ref{a,logamediate}) is converted to the power-law one $a(t)=a_0t^q$, where $q=A$. For the scale factor (\ref{a,logamediate}), the Hubble parameter reads
\begin{equation}
 \label{H,t,logamediate}
 H=\frac{A\lambda\left(\ln t\right)^{\lambda-1}}{t}.
\end{equation}
Furthermore, by using Equations (\ref{{varepsilon}_1}) and (\ref{{varepsilon}_{i+1}}), it is simple to show that the first three slow-roll Hubble parameter are given as follows:
\begin{eqnarray}
 \label{{varepsilon}_1,logamediate}
 \varepsilon_{1} &=& \frac{\left(\ln t-\lambda+1\right)}{A\lambda\left(\ln t\right)^{\lambda}},
 \\
 \label{{varepsilon}_2,logamediate}
 \varepsilon_{2} &=& -\frac{(\lambda-1)\left(\ln t-\lambda\right)}{A\lambda\left(\ln t-\lambda+1\right)\left(\ln t\right)^{\lambda}},
 \\
 \varepsilon_{3} &=& -\frac{\left[\lambda\left(\ln t\right)^{2}-\left(2\lambda^{2}-\lambda+1\right)\ln t+(\lambda-1)\lambda^{2}\right]}{A\lambda\left(\ln t\right)^{\lambda}\left(\ln t-\lambda\right)\left(\ln t-\lambda+1\right)}.
 \nonumber
 \\
 \label{{varepsilon}_3,logamediate}
\end{eqnarray}

It should be noted that unlike the power-law and intermediate DBI inflationary models, here because of the complexity of the calculations, we cannot find the analytical functions for the brane potential $V(\phi)$ and the background warp factor $f(\phi)$ corresponding to the logamediate DBI inflation. However, it is useful to find the asymptotic behaviors of these quantities. To this end, we first simplify Equation (\ref{{dot}{{phi}},H'}) for the logamediate scale factor (\ref{a,logamediate}) and get
\begin{equation}
 \label{{dot}{{phi}},logamediate,exact}
 \dot{\phi}=\frac{\sqrt{2c_{s}A\lambda\left(\ln t\right)^{\lambda-2}\left[\ln t-(\lambda-1)\right]}}{t}.
\end{equation}
Then, we follow the procedure of \citet{Barrow2007}, and consider the late time limit $t\gg1$, allowing us to neglect the term  $(\lambda-1)$ versus $\ln t$ in the above equation. Consequently, we can rewrite the above equation as
\begin{equation}
 \label{{dot}{{phi}},logamediate}
 \dot{\phi}=\frac{\sqrt{2c_{s}A\lambda}\left(\ln t\right)^{\frac{\lambda-1}{2}}}{t}.
\end{equation}
The above differential equation can easily be solved to give the evolutionary behavior of the inflaton field versus time as
\begin{equation}
 \label{{phi},t,logamediate}
 \phi=\frac{2\sqrt{2c_{s}A\lambda}\left(\ln t\right)^{\frac{\lambda+1}{2}}}{\lambda+1},
\end{equation}
where the constant of integration has been set equal to zero without loss of generality. The solution of the above equation for $t$ is
\begin{equation}
 \label{t,{phi},logamediate}
 t=\exp\left[\left(\frac{(\lambda+1)^{2}}{8c_{s}A\lambda}\right)^{\frac{1}{\lambda+1}}\phi^{\frac{2}{\lambda+1}}\right].
\end{equation}
Substituting the above solution into Equation (\ref{H,t,logamediate}), the Hubble parameter is obtained as
\begin{equation}
 \label{H,{phi},logamediate}
 H=H_{0}\phi^{\alpha/2}\exp\left[-B\phi^{\beta}\right],
\end{equation}
where we have defined
\begin{eqnarray}
 \label{H_0,logamediate}
 H_{0} &\equiv& A\lambda B^{\lambda-1},
 \\
 \label{B,logamediate}
 B &\equiv& \left(\frac{(\lambda+1)^{2}}{8c_{s}A\lambda}\right)^{\frac{1}{\lambda+1}},
 \\
 \label{{alpha},logamediate}
 \alpha &\equiv& \frac{4(\lambda-1)}{\lambda+1},
 \\
 \label{{beta},logamediate}
 \beta &\equiv& \frac{2}{\lambda+1}.
\end{eqnarray}
For the constant sound speed, from Equation (\ref{c_s}) and using $X=\dot{\phi}^2/2$, we can derive the warp factor of this model as
\begin{equation}
 \label{f,{phi},logamediate}
f(\phi)=f_{0}\phi^{-\alpha/2}\exp\left[2B\phi^{\beta}\right],
\end{equation}
where
\begin{equation}
 \label{f_0,logamediate}
 f_{0}\equiv\frac{1-c_{s}^{2}}{2c_{s}H_{0}}.
\end{equation}
In addition to the late time limit, if we assume the slow-roll conditions, then using Equation (\ref{H,{phi},logamediate}) in (\ref{H,V}) one can obtain the inflationary potential as
\begin{equation}
 \label{V,{phi},logamediate}
 V(\phi)=V_{0}\phi^{\alpha}\exp\left[-2B\phi^{\beta}\right],
\end{equation}
where the parameter $V_0$ is defined as
\begin{equation}
 \label{V_0,logamediate}
 V_{0}\equiv3H_{0}^{2}=3\left(A\lambda B^{\lambda-1}\right)^{2}.
\end{equation}
It can be readily checked that in the case of $c_s=1$, the DBI potential (\ref{V,{phi},logamediate}) reduces to the one obtained for the logamediate inflation in the standard canonical scenario \citep{Barrow2007, Rezazadeh2017, Barrow1995}.

In what follows, we proceed to estimate the inflationary observables for the present model. Using Equations (\ref{H,t,logamediate}) and (\ref{{varepsilon}_1,logamediate}), the power spectrum of scalar density perturbations (\ref{{mathcal}{P}_s}) yields
\begin{equation}
\label{{mathcal}{P}_s,t,logamediate}
{\cal P}_{s}=\frac{\left(A\lambda\right)^{3}\left(\ln t\right)^{3\lambda-2}}{8\pi^{2}c_{s}t^{2}\left(\ln t-\lambda+1\right)}.
\end{equation}
Also, by replacing Equations (\ref{{varepsilon}_1,logamediate})--(\ref{{varepsilon}_3,logamediate}) and $\varepsilon_{s1}=\varepsilon_{s2}=0$ into Equations (\ref{n_s}), (\ref{dn_s/d{ln}k}), and (\ref{r}), we obtain
\begin{align}
 n_{s}=&\left[A\lambda\left(\ln t\right)^{\lambda}\left(\ln t-\lambda+1\right)\right]^{-1}
 \nonumber
 \\
 &\times\Big[A\lambda\left(\ln t\right)^{\lambda+1}-A\lambda(\lambda-1)\left(\ln t\right)^{\lambda}
 \nonumber
 \\
 &-2\left(\ln t\right)^{2}+5(\lambda-1)\ln t-3\lambda^{2}+5\lambda-2\Big],
 \label{n_s,logamediate}
 \\
 \frac{dn_{s}}{d\ln k}= &(\lambda-1)\left[A\lambda\left(\ln t\right)^{\lambda}\left(\ln t-\lambda+1\right)\right]^{-2}
 \nonumber
 \\
 &\times\Big[2\left(\ln t\right)^{3}-(7\lambda-4)\left(\ln t\right)^{2}
 \nonumber
 \\
 &+\left(8\lambda^{2}-9\lambda+3\right)\left(\ln t\right)-\lambda\left(3\lambda^{2}-5\lambda+2\right)\Big],
 \label{dn_s/d{ln}k,logamediate}
 \\
 r = &\frac{16c_{s}\left(\ln t-\lambda+1\right)}{A\lambda\left(\ln t\right)^{\lambda}}.
 \label{r,logamediate}
\end{align}
Notice that for the case of $\lambda=1$, the above equations reduce to Equations (\ref{n_s,power-law}), (\ref{dn_s/d{ln}k,power-law}), and (\ref{r,power-law}) obtained in Sec. \ref{section:power-law} for the DBI power-law inflation. It makes sense because in this case the logamediate scale factor (\ref{a,logamediate}) changes into the power-law form $a(t)\propto t^{q}$ in which $q=A$.

Solving the differential equation (\ref{dN}) with the Hubble parameter (\ref{H,t,logamediate}), one can find
\begin{equation}
 \label{t,N,logamediate}
 t=\exp\left[\left(\left(\ln t_{e}\right)^{\lambda}-\frac{N}{A}\right)^{\frac{1}{\lambda}}\right],
\end{equation}
where the integration constant has been determined by considering the end of inflation condition, i.e. $N_e\equiv N(t=t_e)=0$. In order to find the parameter $t_e$ in terms of the other model parameters, we fix the amplitude of scalar perturbations ${\cal P}_{s}$ at the horizon exit $e$-fold number $N_{*}$, according to the Planck observational results. To this aim, we first obtain $\mathcal{P}_{s*}$ by replacing Equation (\ref{t,N,logamediate}) into (\ref{{mathcal}{P}_s,t,logamediate}) as
\begin{align}
 \mathcal{P}_{s*}= &\frac{\left(A\lambda\right)^{3}\left(\left(\ln t_{e}\right)^{\lambda}-\frac{N_{*}}{A}\right)^{\frac{3\lambda-2}{\lambda}}}{8\pi^{2}c_{s}\left[\left(\left(\ln t_{e}\right)^{\lambda}-\frac{N_{*}}{A}\right)^{1/\lambda}-\lambda+1\right]}
 \nonumber
 \\
 &\times\exp\left[-2\left(\left(\ln t_{e}\right)^{\lambda}-\frac{N_{*}}{A}\right)^{1/\lambda}\right].
 \label{{mathcal}{P}_s,N_*,logamediate}
\end{align}
We recall that the quantity $\mathcal{P}_{s*}$ is a constant number constrained as $\ln\left[10^{10}\mathcal{P}_{s*}\right]=3.094\pm0.034$ due to the Planck 2015 TT,TE,EE+lowP data at 68\% CL \citep{Planck2015}. We apply this constraint to obtain the parameter $t_e$ in terms of the other model parameters. It should be noted that Equation (\ref{{mathcal}{P}_s,N_*,logamediate}) cannot be solved analytically, and we have to employ a numerical method for this purpose. In the next step, we replace $t_e$ from Equation (\ref{{mathcal}{P}_s,N_*,logamediate}) into (\ref{t,N,logamediate}) to find the time of horizon crossing $t_*$ for given parameters $n$, $A$, $\lambda$, and $N_*$. Surprisingly, our numerical computations imply that $t_*$ is completely independent of $N_*$. Such a situation also appears in the study of logamediate inflation in $f(T)$-gravity \citep{Rezazadeh2017}. Here, we repeat the deduction of \citet{Rezazadeh2017} to justify the mentioned surprising consequence. Following \citet{Rezazadeh2017}, we calculate the partial derivative of Equation (\ref{{mathcal}{P}_s,N_*,logamediate}) with respect to $N_*$, which leads to
\begin{equation}
 \label{dt_e}
 \frac{\partial t_{e}}{\partial N_{*}}=\frac{t_{e}}{A\lambda\left(\ln t_{e}\right)^{\lambda-1}},
\end{equation}
where we have taken into account $\partial\mathcal{P}_{s*}/\partial N_{*}=0$ in the above derivation. Also, taking the partial derivative of Equation (\ref{t,N,logamediate}) with respect to $N_*$, we attain
\begin{align}
 \frac{\partial t_{*}}{\partial N_{*}}= &\frac{\exp\left[\left(\left(\ln t_{e}\right)^{\lambda}-\frac{N_{*}}{A}\right)^{1/\lambda}\right]}{\left(\left(\ln t_{e}\right)^{\lambda}-\frac{N_{*}}{A}\right)^{(\lambda-1)/\lambda}}
 \nonumber
 \\
 &\times\left(\frac{\left(\ln t_{e}\right)^{\lambda-1}}{t_{e}}\frac{\partial t_{e}}{\partial N_{*}}-\frac{1}{A\lambda}\right).
 \label{dt_*}
\end{align}
Now, it is evident that inserting Equation (\ref{dt_e}) into the above equation gives rise to $\partial t_{*}/\partial N_{*}=0$, implying that the time horizon exit does not depend on $N_*$, at all. Consequently, the inflationary observables $n_s$, $dn_s/d\ln k$, and $r$ estimated at $t_*$, become independent of $N_*$. Therefore, the inflationary observables in our model depend on the three free parameters $A$, $\lambda$, and $c_s$.

In Figure \ref{figure:r,n_s,logamediate}, using Equations (\ref{n_s,logamediate}) and (\ref{r,logamediate}) we plot the $r-n_s$ diagram for the logamediate inflation (\ref{a,logamediate}) driven by the DBI scalar field (\ref{{mathcal}{L}}) for different values of $c_s$ and $\lambda$ with varying $A$ in the range of $A>0$. Also, the joint regions 68\% and 95\% CL allowed by Planck 2015 data have been marginalized in the figure. It is clear from Figure \ref{figure:r,n_s,logamediate} that for the case $c_s=1$ associated with the logamediate inflation in the standard canonical setting, the $r-n_s$ diagram is not consistent with the Planck 2015 data. But in the DBI scenario with $c_s=0.1$, for instance, the prediction of logamediate inflation can lie inside the 68\% CL region of the Planck 2015 TT,TE,EE+lowP data \citep{Planck2015}. It is evident from the figure that in order for our model to be consistent with 68\% CL (95\% CL) bounds of these data, the parameter $\lambda$ has to be chosen in the range $\lambda \lesssim 8.9$ ($\lambda \lesssim 9.39$).

\begin{figure*}
\begin{center}
\scalebox{1}[1]{\includegraphics{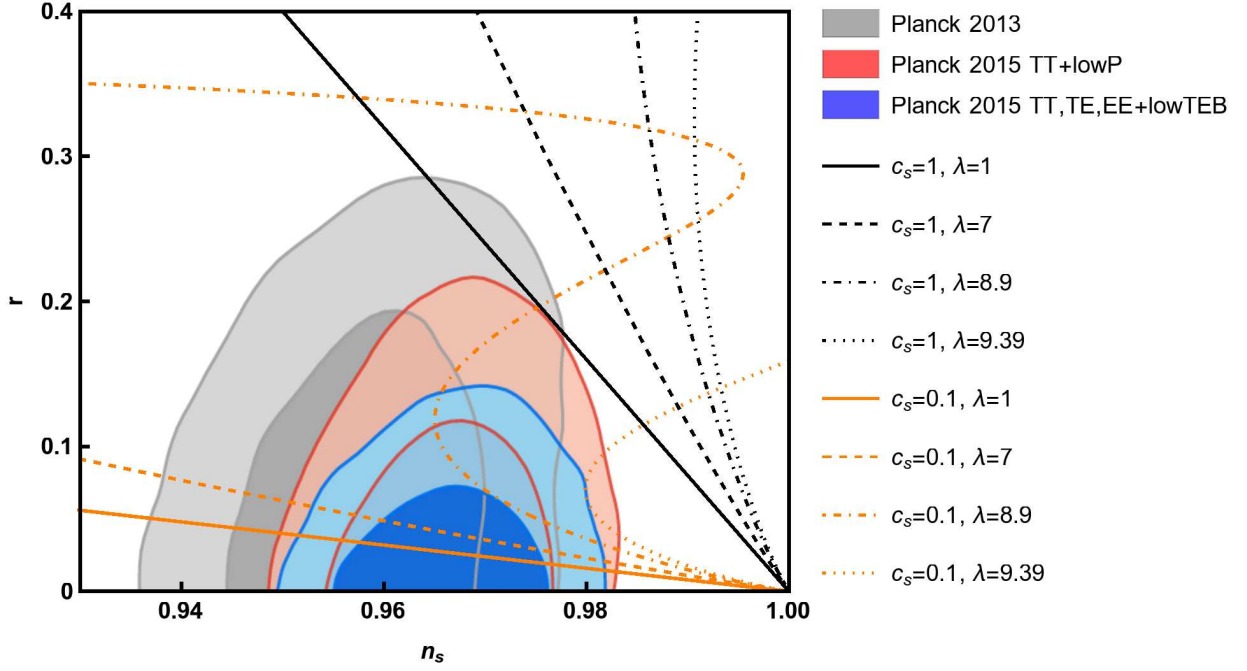}}
\caption{Same as Figure \ref{figure:r,n_s,power-law}, but for the logamediate inflation (\ref{a,logamediate}) in the DBI setting (\ref{{mathcal}{L}}) for different values of $c_s$ and $\lambda$ with varying $A$ in the range of $A>0$.
}
\label{figure:r,n_s,logamediate}
\end{center}
\end{figure*}

\begin{table*}
  \centering
  \caption{Same as Table \ref{table:intermediate}, but for the DBI logamediate inflation with $c_s=0.1$ and different values of $\lambda$.}
  \scalebox{0.8}{
  \begin{tabular}{ccccc}
  \hline
  \hline
  $\quad$  $\lambda$ $\quad$ & $\quad\quad\quad$ $A$ (68\% CL) $\quad\quad\quad$ & $\quad\quad\quad$ $\frac{dn_s}{d\ln k}$ $\quad\quad\quad$ & $\quad\quad\quad$ $A$ (95\% CL) $\quad\quad\quad$ & $\quad\quad\quad$ $\frac{dn_s}{d\ln k}$ $\quad\quad\quad$ \\
  \hline
  $1$ & $[47, 83]$ & $0$ & $[41, 111]$ & $0$\\
  $2$ & $[1.45, 2.4]$ & $[1.922 \times 10^{-5} , 6.147 \times 10^{-5}]$ & $[1.29, 3.13]$ & $[1.045 \times 10^{-5} , 8.061 \times 10^{-5}]$ \\
  $3$ & $[0.061, 0.095]$ & $[4.201 \times 10^{-5}, 1.316 \times 10^{-4}]$ & $[0.0548, 0.122]$ & $[2.219\times 10^{-5} , 1.739 \times 10^{-4}]$ \\
  $4$ & $[2.9 \times 10^{-3}, 4.2 \times 10^{-3}]$ & $[7.198 \times 10^{-5} , 2.126 \times 10^{-4}]$ & $[2.60 \times 10^{-3}, 5.27 \times 10^{-3}]$ & $[3.748 \times 10^{-5}, 2.939 \times 10^{-4}]$ \\
  $5$ & $[1.5 \times 10^{-4}, 2.0 \times 10^{-4}]$ & $[1.105 \times 10^{-4} , 2.923 \times 10^{-4}]$ & $[1.36 \times 10^{-4}, 2.50 \times 10^{-4}]$ & $[5.294 \times 10^{-5} , 4.098 \times 10^{-4}]$ \\
  $6$ & $[7.55 \times 10^{-6}, 9.96 \times 10^{-6}]$ & $[1.649 \times 10^{-4}, 5.049 \times 10^{-4}]$ & $[6.99 \times 10^{-6}, 1.18 \times 10^{-5}]$ & $[8.660 \times 10^{-5}, 6.966 \times 10^{-4}]$ \\
  $7$ & $[4.15 \times 10^{-7}, 5.14 \times 10^{-7}]$ & $[2.463 \times 10^{-4}, 7.205 \times 10^{-4}]$ & $[3.88 \times 10^{-7}, 5.93 \times 10^{-7}]$ & $[1.248 \times 10^{-4}, 1.027 \times 10^{-3}]$ \\
  $8$ & $[2.36 \times 10^{-8}, 2.73 \times 10^{-8}]$ & $[3.830 \times 10^{-4}, 9.970 \times 10^{-4}]$ & $[2.23 \times 10^{-8}, 3.07 \times 10^{-8}]$ & $[1.864 \times 10^{-4}, 1.478 \times 10^{-3}]$ \\
  $9$ & --- & --- & $[1.35 \times 10^{-9}, 1.62 \times 10^{-9}]$ & $[-3.351 \times 10^{-3}, 4.675 \times 10^{-4}]$ \\
  $\lambda \gtrsim 9.39$ & --- & --- & --- & --- \\
  \hline
\end{tabular}
}
\label{table:logamediate}
\end{table*}

\begin{table*}
  \centering
  \caption{Estimated values of inflationary observables for the DBI logamediate inflation with $\lambda=2$ and some typical values of $c_s$ and $A$.}
\scalebox{0.8}{
\begin{tabular}{ccccccc}
  \hline
  \hline
  $\qquad$  $c_s$ $\qquad$ & $\qquad$ $A$ $\qquad$ & $\qquad$ $n_s$ $\qquad$ & $\frac{dn_s}{d\ln k}$ & $\qquad$ $r$ $\qquad$ & $\qquad$ $f_{{\rm NL}}^{{\rm DBI}}$ $\qquad$ & $r-n_s$ Consistency \\
  \hline
  $0.1$ & $2$ & $0.9701$ & $2.925 \times 10^{-5}$ & $0.0247$ & $-32.083$ & 68\% CL \\
  $0.2$ & $2$ & $0.9694$ & $3.141 \times 10^{-5}$ & $0.0506$ & $-7.778$ & 68\% CL \\
  $0.3$ & $2$ & $0.9690$ & $3.277 \times 10^{-5}$ & $0.0770$ & $-3.277$ & 95\% CL \\
  $0.4$ & $2$ & $0.9687$ & $3.379 \times 10^{-5}$ & $0.1037$ & $-1.701$ & 95\% CL \\
  $0.5$ & $2$ & $0.9685$ & $3.461 \times 10^{-5}$ & $0.1307$ & $-0.972$ & 95\% CL \\
  $0.6$ & $2.3$ & $0.9728$ & $2.551 \times 10^{-5}$ & $0.1352$ & $-0.576$ & 95\% CL \\
  $0.7$ & $2$ & $0.9681$ & $3.590 \times 10^{-5}$ & $0.1852$ & $-0.337$ & --- \\
  \hline
\end{tabular}
}
\label{table:logamediate,observables}
\end{table*}

In Table \ref{table:logamediate}, we present the ranges of parameter $A$ for which the result of our model with $c_s=0.1$ and for different values of $\lambda$, is compatible with the the Planck 2015 TT,TE,EE+lowP data \citep{Planck2015}. The ranges of $A$ in this table have been obtained in a similar method applied in Table \ref{table:intermediate} prepared for the intermediate DBI inflation. Also, in this table, we have estimated the running of the scalar spectral index $dn_s/d\ln k$ in our model by using Equation (\ref{dn_s/d{ln}k,logamediate}). The predicted values for this observable satisfy the 95\% CL constraint of the Planck 2015 TT,TE,EE+lowP data \citep{Planck2015}.

In Figure \ref{figure:A,lambda,logamediate}, the parameter space of $A$ and $\lambda$ are specified for some typical values of $c_s$, in which the result of the $r-n_s$ diagram is compatible with the 68\% or 95\% CL joint regions of the Planck 2015 TT,TE,EE+lowP data \citep{Planck2015}. To plot this diagram, we have utilized the same numerical method as that used in Figure \ref{figure:A,lambda,intermediate} corresponding to the intermediate DBI inflation.

\begin{figure*}
\begin{minipage}[b]{1\textwidth}
\subfigure{ \includegraphics[width=.48\textwidth]%
{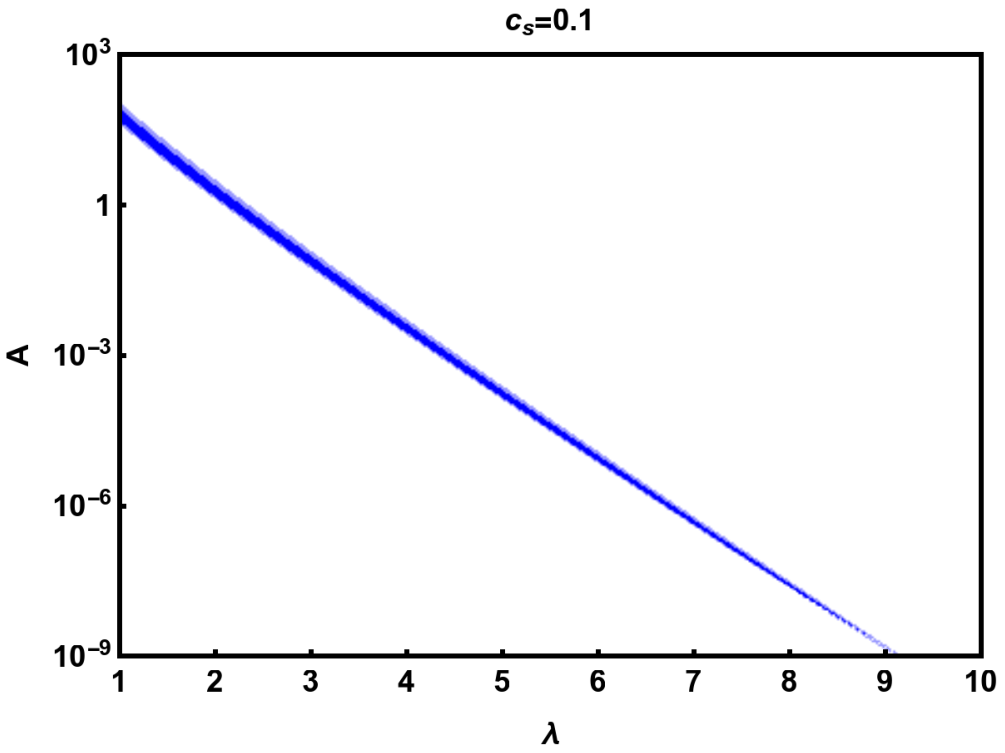}} \hspace{.1cm}
\subfigure{ \includegraphics[width=.48\textwidth]%
{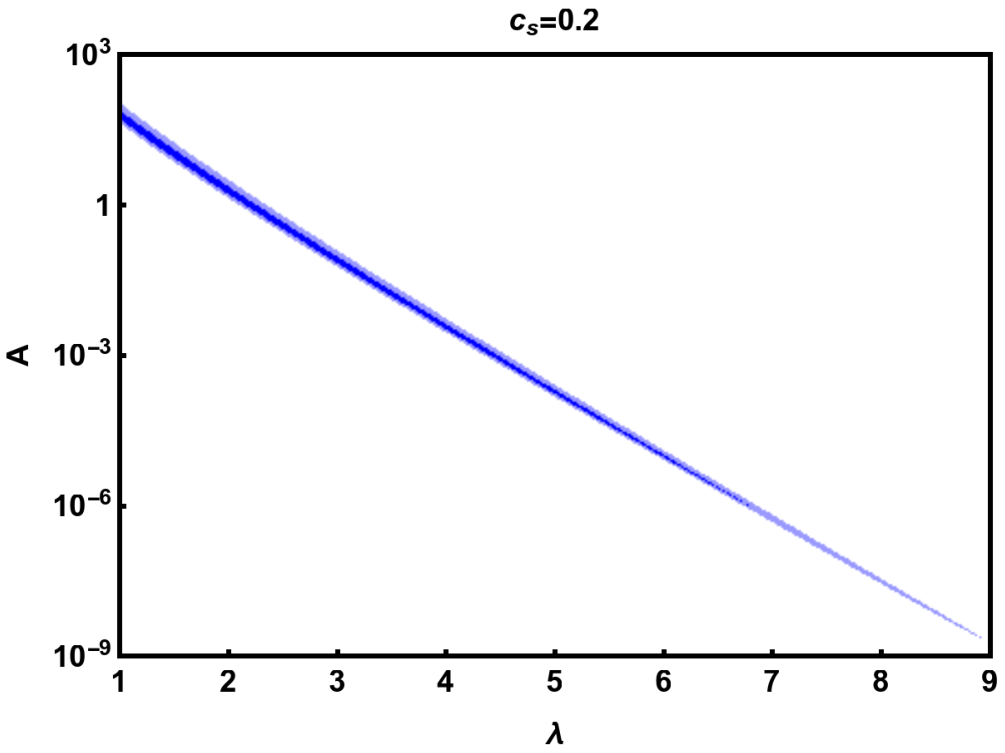}}
\end{minipage}
\begin{minipage}[b]{1\textwidth}
\subfigure{ \includegraphics[width=.48\textwidth]%
{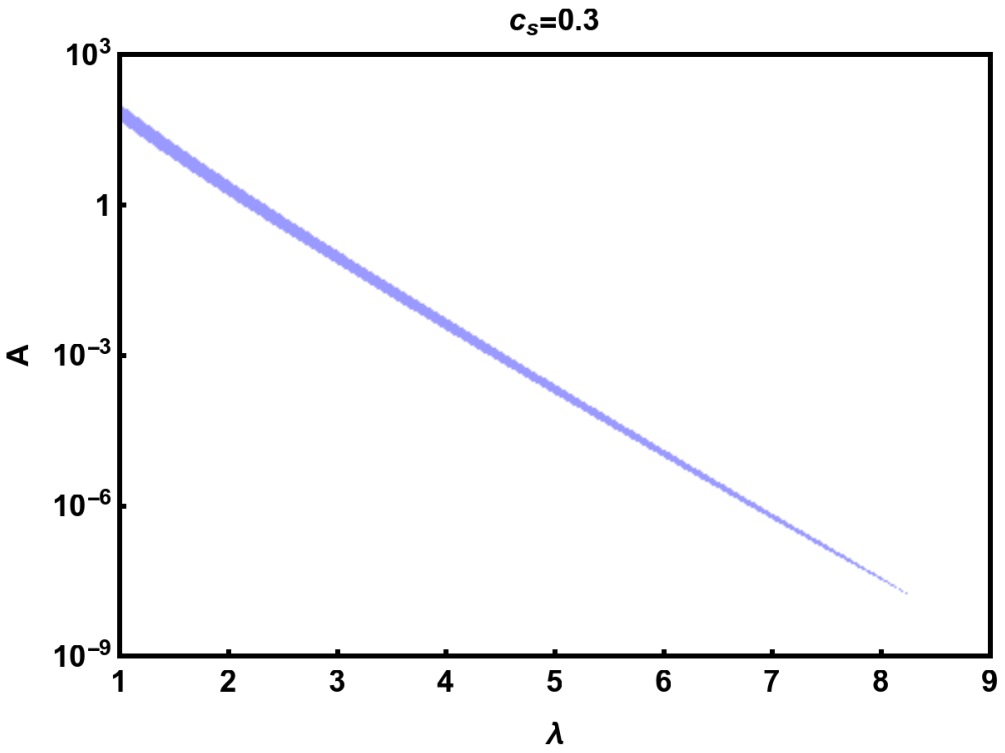}} \hspace{.1cm}
\subfigure{ \includegraphics[width=.48\textwidth]%
{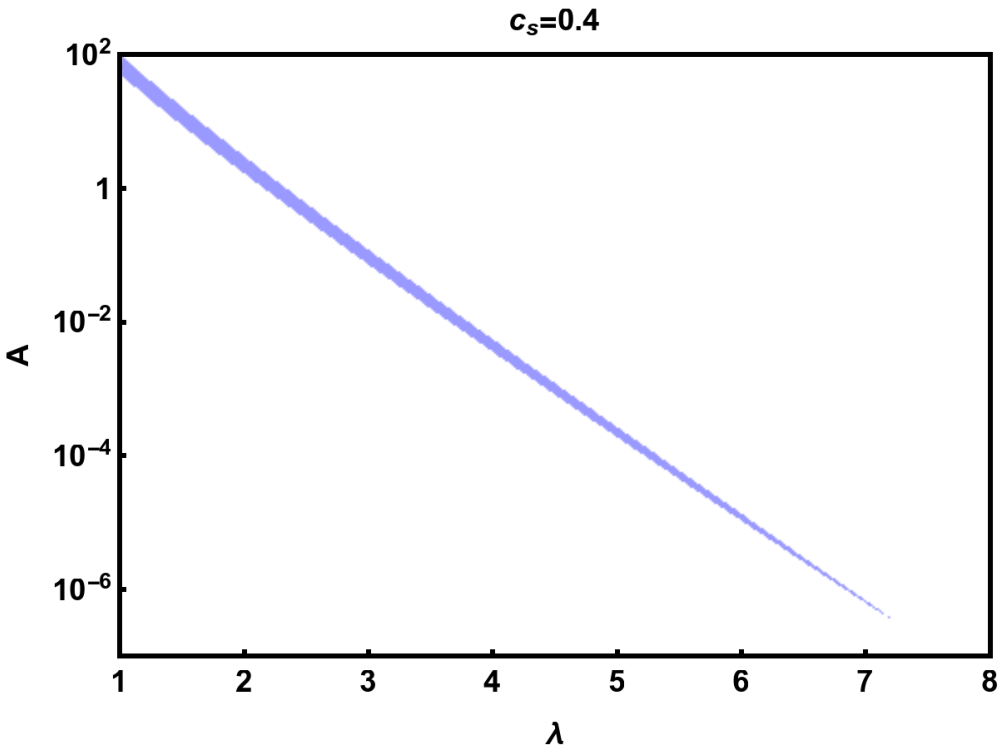}}
\end{minipage}
\begin{minipage}[b]{1\textwidth}
\subfigure{ \includegraphics[width=.48\textwidth]%
{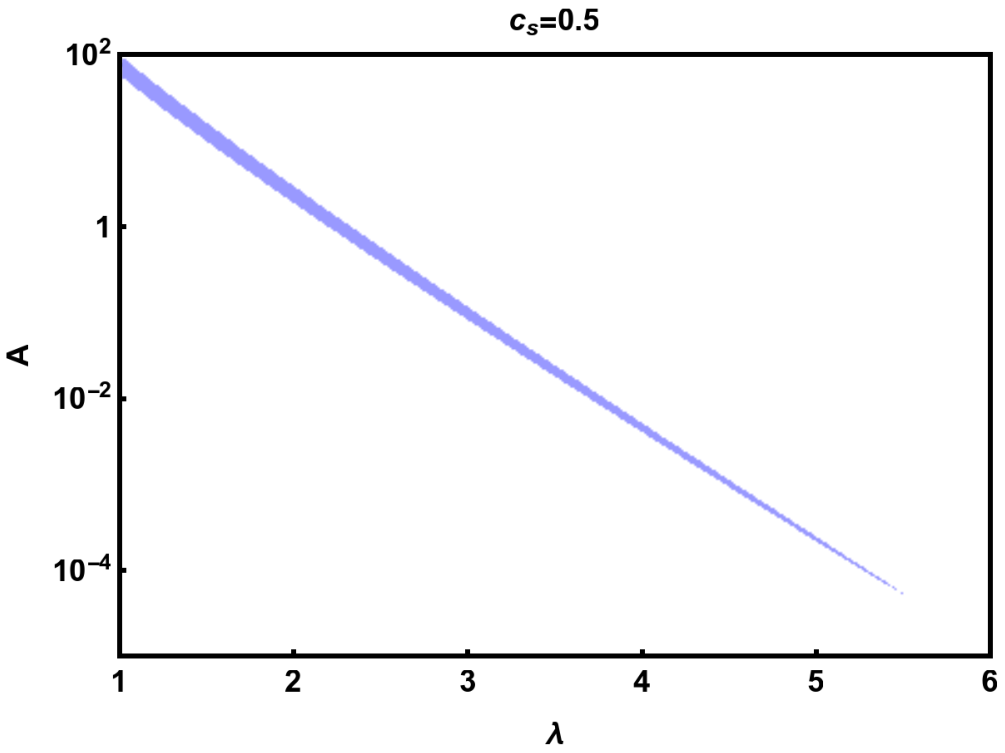}} \hspace{.1cm}
\subfigure{ \includegraphics[width=.48\textwidth]%
{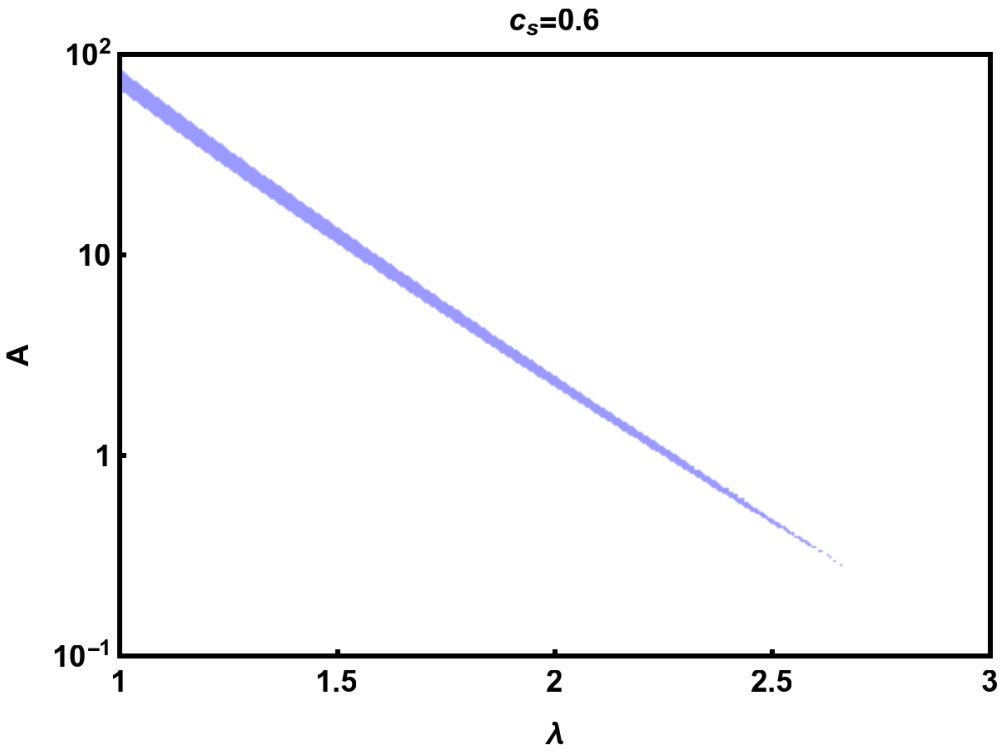}}
\end{minipage}
\caption{Same as Figure \ref{figure:A,lambda,intermediate}, but for the logamediate inflation (\ref{a,logamediate}) in the DBI scenario (\ref{{mathcal}{L}}) with different values of the constant sound speed $c_s$.}
\label{figure:A,lambda,logamediate}
\end{figure*}

At the end of this section, we present some numerical values for the inflationary observables in the logamediate DBI inflation. These values are listed in Table \ref{table:logamediate,observables} for $\lambda=2$ and some typical values of $c_s$ and $A$. For each case we have examined the $r-n_s$ consistency and concluded that our results are in agreement with those illustrated in Figure \ref{figure:A,lambda,logamediate}.

\section{Conclusions}
\label{section:conclusions}

Here, we studied the power-law, intermediate, and logamediate inflations driven by the DBI non-canonical scalar field. The DBI inflation is inspired from string theory, and in this model the role of inflaton is performed by the radial coordinate of a D3-brane moving on the throat of the compactification space. We first investigated the basic equations governing the cosmological background evolution in the DBI inflationary scenario. In particular, we showed that the slow-roll first Friedmann equation in the DBI setting is identical to that of the standard canonical scenario. Subsequently, with the help of scalar and tensor power spectra in the setup of DBI inflation, we obtained the inflationary observables including the scalar spectral index $n_s$, the running of the scalar spectral index $dn_s/d\ln k$, and the tensor-to-scalar ratio $r$. We showed that the 95\% CL constraint from temperature and polarization data of the Planck 2015 collaboration on the non-Gaussianity parameter leads to the lower bound $c_{s}\geq0.087$ on the sound speed of the DBI inflation.

In addition, we assumed the sound speed to be constant during inflation and examined viability of the power-law, intermediate, and logamediate DBI inflationary models in light of the Planck 2015 observational data. The first scale factor that we investigated was the power-law scale factor $a(t)\propto t^{q}$ where $q>1$. We showed that in the slow-roll approximation, this scale factor arises from the ascending exponential warp factor $f(\phi)\propto\exp\left(\sqrt{\frac{2}{c_{s}q}}\,\phi\right)$ and the descending exponential potential $V(\phi)\propto\exp\left(-\sqrt{\frac{2}{c_{s}q}}\,\phi\right)$. We further showed that the intermediate scale factor $a(t)\propto\exp\big(At^{\lambda}\big)$ where $A>0$ and $0<\lambda<1$, is driven by the power-law warp factor $f(\phi)\propto\phi^{2\left(2-\lambda\right)/\lambda}$ and the inverse power-law potential $V(\phi)\propto\phi^{-4(1-\lambda)/\lambda}$. In the study of logamediate scale factor $a(t)\propto\exp\Big[A\left(\ln t\right)^{\lambda}\Big]$ where $A>0$ and $\lambda\geq1$, we derived the warp factor $f(\phi)\propto\phi^{-\alpha/2}\exp\left[2B\phi^{\beta}\right]$ and the inflationary potential $V(\phi)\propto\phi^{\alpha}\exp\left[-2B\phi^{\beta}\right]$ where $\alpha$, $\beta$, and $B$ are positive constants. Note that in the case of $c_s=1$, the slow-roll results of the DBI inflationary potentials corresponding to the aforementioned scale factors are the same as those obtained in the standard canonical framework.

Our study implies that, although the power-law, intermediate, and logamediate inflationary models in the standard canonical framework are completely ruled out by the Planck 2015 results, in the DBI setting with constant sound speed they can be compatible with the Planck observations. We showed that the predictions of these DBI models in the $r-n_s$ plane can lie inside the 68\% CL joint region of the Planck 2015 TT,TE,EE+lowP data \citep{Planck2015}, if we take the sound speed $c_s$ to be sufficiently less than the light speed ($c=1$). For the power-law, intermediate, and logamediate DBI models, we specified their parameter space for which the $r-n_s$ predictions of these models are compatible with 68\% or 95\% CL marginalized joint regions of the Planck 2015 TT,TE,EE+lowP data \citep{Planck2015}. In our work, we further estimated the running of the scalar spectral index $dn_s/d\ln k$ for the allowed ranges of the parameter space of the intermediate and logamediate DBI inflationary models and concluded that the estimated values satisfy the 95\% CL constraint of the Planck 2015 TT,TE,EE+lowP data \citep{Planck2015}.

\acknowledgments

The authors thank the referee for valuable comments.

\end{document}